\documentclass[a4paper,12pt]{article}

\usepackage{amsmath}
\usepackage{amssymb}
\usepackage{bbm}
\usepackage{graphicx}
\usepackage{accents}
\usepackage{amsthm}
\usepackage{color}
\usepackage{graphics}
\usepackage{epsfig}
\usepackage{cancel}
\usepackage{epstopdf}
\usepackage{hyperref}
\usepackage{xcolor}
\hypersetup{
colorlinks,
linkcolor={blue!50!black},
citecolor={blue!50!black},
urlcolor={blue!80!black}
}
\usepackage{natbib}
\usepackage{stmaryrd} 
\usepackage[top = 1in, left = 0.5in, bottom = 1in, right=0.5in]{geometry}

\newcommand{\mbf}[1]{\mathbf{#1}}
\newcommand{\bbm}[1]{\mathbbm{#1}}
\newcommand{\mcal}[1]{\mathcal{#1}}
\newcommand{\bsym}[1]{\boldsymbol{#1}}
\newcommand{\bbbm}[1]{\bsym{\bbm{#1}}}

\newcommand{\jump}[1]{\llbracket {#1} \rrbracket}
\newcommand{\wt}[1]{\widetilde{#1}}
\newcommand{\wh}[1]{\widehat{#1}}
\newcommand{\pd}[2]{\frac{\partial #1}{\partial #2}}

\DeclareMathOperator{\CGright}{\bsym{C}}
\DeclareMathOperator{\Dgrad}{\bsym{F}}

\DeclareMathOperator{\bod}{\mcal{B}}
\DeclareMathOperator{\bbod}{\partial \mcal{B}} 
\DeclareMathOperator{\strPK}{\mbf{P}} 
\DeclareMathOperator{\strC}{\bsym{\sigma}} 
\DeclareMathOperator{\vol}{\mcal{V}}

\DeclareMathOperator{\elecR}{\bbbm{E}} 
\DeclareMathOperator{\elecC}{\bbbm{e}} 
\DeclareMathOperator{\disR}{\bbbm{D}} 
\DeclareMathOperator{\disC}{\bbbm{d}} 
\DeclareMathOperator{\polR}{\bbbm{P}} 
\DeclareMathOperator{\polC}{\bbbm{p}} 
\DeclareMathOperator{\potR}{\bbbm{A}} 
\DeclareMathOperator{\potC}{\bbbm{a}} 

\DeclareMathOperator{\Div}{\text{Div}}
\DeclareMathOperator{\Grad}{\text{Grad}}
\DeclareMathOperator{\grad}{\text{grad}}
\DeclareMathOperator{\Curl}{\text{Curl}}
\DeclareMathOperator{\curl}{\text{curl}}

\DeclareMathOperator{\pot}{\mbf{A}} 

\DeclareRobustCommand{\rchi}{{\mathpalette\irchi\relax}}
\newcommand{\irchi}[2]{\raisebox{\depth}{$#1\chi$}} 

\usepackage{mdframed} 
\newmdtheoremenv[%
linecolor=white,
leftmargin=0,%
rightmargin=0,
backgroundcolor=white!10,%
innertopmargin=0pt,%
ntheorem]{sidenote}{Remark}[section]

\usepackage{array}
\newcolumntype{L}[1]{>{\raggedright\let\newline\\\arraybackslash\hspace{0pt}}m{#1}}
\newcolumntype{C}[1]{>{\centering\let\newline\\\arraybackslash\hspace{0pt}}m{#1}}
\newcolumntype{R}[1]{>{\raggedleft\let\newline\\\arraybackslash\hspace{0pt}}m{#1}}

\usepackage{tabu}

\title{On equilibrium equations and their perturbations using three different variational formulations of nonlinear electroelastostatics} 

\usepackage{fancyhdr}
\makeatletter
\let\runauthor\@author
\let\runtitle\@title
\makeatother
\author{Prashant Saxena\thanks{Glagsow Computational Engineering Centre, James Watt School of Engineering, University of Glasgow, Glasgow G12 8LT, UK. Email: prashant.saxena@glasgow.ac.uk} \and Basant Lal Sharma\thanks{Department of Mechanical Engineering, Indian Institute of Technology Kanpur, Kanpur, U. P. 208016, India. Email: bls@iitk.ac.in}}

\author{Prashant Saxena$^1$\thanks{Corresponding author email: prashant.saxena@glasgow.ac.uk}, Basant Lal Sharma$^2$ \\[2ex]
{\small $^1$Glagsow Computational Engineering Centre, James Watt School of Engineering}\\ {\small University of Glasgow, Glasgow G12 8LT, UK}\\
{\small $^2$Department of Mechanical Engineering, Indian Institute of Technology Kanpur}\\ {\small Kanpur, Uttar Pradesh 208016, India}}

\date{\small This document was last updated on \today.}
\date{}

\allowdisplaybreaks 

\begin{document}

\maketitle

\begin{abstract}
We derive the equations of nonlinear electroelastostatics using three different variational formulations involving the deformation function and an independent field variable representing the electric character -- considering either one of the electric field $\elecR$, electric displacement $\disR$, or electric polarization $\polR$.
The first variation of the energy functional results in the set of Euler-Lagrange partial differential equations which are the equilibrium equations, boundary conditions, { and certain constitutive equations} for the electroelastic system.
The partial differential equations for obtaining the bifurcation point have been also found using the second variation based bilinear functional.
We show that the well-known Maxwell stress in vacuum is a natural outcome of the derivation of equations from the variational principles and does not depend on the formulation used.
As a result of careful analysis it is found that there are certain terms in the bifurcation equation which appear difficult to obtain by an ordinary perturbation based analysis of the Euler-Lagrange equation.
From a practical viewpoint, the formulations based on $\elecR$ and $\disR$ result in simpler equations and are anticipated to be more suitable for analysing problems of stability as well as post-buckling behaviour. 

\end{abstract}


\section*{Introduction}

Early development of a nonlinear theory of elastic dielectrics is attributed to the seminal work of \cite{Toupin1956}.
Past couple of decades have seen a surge in the study of a nonlinear theory of electroelasticity within the framework of Continuum Mechanics \citep{Dorfmann2005a, McMeeking2005} largely motivated by the development of electro-active polymers (EAPs).
EAPs are capable of producing large deformations in the presence of electric fields and alternatively can be used to convert mechanical deformation to electric potential difference \citep{Pelrine2000a, Kofod2001, Jung2008}.
Their use has been demonstrated in the development of artifical muscles and robotic manipultors \citep{Wingert2006, Shintake2016}, haptic interfaces \citep{Ozsecen2010}, electric generators \citep{Pelrine2001}, propulsion systems \citep{Michel2008} and sensing equipments \citep{O'Halloran2008}.

Development of variational principles for nonlinear electromechanics is essential to derive a consistent set of partial differential equations and suitable boundary conditions, analysis of stability of equilibrium, and to perform computations based on the finite element method.
Initial variational principles have been provided by \cite{Pak1986} and \cite{Yang1995}.
\cite{McMeeking2005} constructed principle of virtual work using the displacement and the electrostatic potential as the independent variable.
\cite{Ericksen2007a} has revisited the theory of Toupin from a 
different starting point and derived a variational principle.
Variational principles that are applicable up to derivation of equilibrium equations by taking the first variation are also presented by \cite{Bustamante2009} and
they are discussed in more detail later in the book by \cite{Dorfmann2014b}.
A variational principle with application to numerical computations (albeit only for computing equilibrium) has been presented by \cite{Vu2012a}.

In this work, we present variational formulations of electroelasticity considering each one of either the electric field $\elecR$, electric displacement $\disR$, or the electric polarization $\polR$ as the independent variable.
For the formulations with $\elecR$ and $\disR$, we start with the known potential energy functional given by \cite{Dorfmann2014b}, while for the formulation with $\polR$ as the independent variable, we consider the potential energy functional used by \cite{Liu2014}. 
First variation of the functional gives the equilibrium equations and boundary conditions while second variation gives the equations for the critical point corresponding to a bifurcation of solution (onset of instability)  {\citep{Koiter2,Koiter,Heijden, Hill1957}}.
 { Critical instability points have also been studied by a direct perturbation of governing equations by \cite{Bertoldi2011, Dorfmann2014a}
}
We calculate the same Maxwell stress tensor outside the body for each of these {three} formulations  { \citep{Bustamante2009b}}.

Computation of first variation for formulations based on $\disR$ apriori requires the Maxwell's law that $\disR$ should satisfy.
The Maxwell's law corresponding to its conjugate vector $\elecR$ is an outcome of the process along with a constitutive relationship between $\elecR$ and $\disR$ via the energy density function.
Similar observation is made for the variational formulation based on $\elecR$.
Differently from the above two, computation of first variation for formulations based on $\polR$ apriori requires the Maxwell's laws both for $\elecR$ and $\disR$.
It is found that the formulations based on $\elecR$ and $\disR$ result in simpler equations and are more amenable to the theory of `total energy' and `total stress' developed by \cite{Dorfmann2006} as the first Piola--Kirchhoff stress is obtained, quite simply, as the derivative of the energy density function with respect to the deformation gradient.
However, this is not the case for {the formulation} based on $\polR$ primarily because polarisation vanishes outside the body.

This paper is organised as follows.
After briefly introducing the mathematical preliminaries, 
in Section \ref{sec: basic eqns} we introduce the system under study and present the basic equations of nonlinear electroelastostatics.
In Sections \ref{sec: D formulation} and \ref{sec: E formulation}, we present the derivations of first and second variations of the potential energy functionals corresponding to $\disR$ and $\elecR$, respectively.
In Section \ref{sec: P formulation}, we present the first variation of the potential energy functional corresponding to $\polR$ and then derive the equations for critical point by linearising the equilibrium equations.
Some detailed calculations are presented in the three appendices.

\subsection*{Mathematical preliminaries}
\label{sec: notation}

{
Direct notation of tensor algebra and tensor calculus is adopted throughout.
The scalar product of two vectors $\mbf{a}$ and $\mbf{b}$ is denoted as $\mbf{a} \cdot \mbf{b} = [\mbf{a}]_i [\mbf{b}]_i$ where a repeated index implies summation according to Einstein's summation convention.
 {The vector (cross) product of two vectors $\mbf{a}$ and $\mbf{b}$ is denoted as $\mbf{a} \wedge \mbf{b}$ with $[\mbf{a} \wedge \mbf{b}]_i = \varepsilon_{ijk} [\mbf{a}]_j [\mbf{b}]_k $, $\varepsilon_{ijk}$ being the permutation symbol.}
The tensor product of two  vectors $\mbf{a}$ and $\mbf{b}$ is a second order tensor $\mbf{H} = \mbf{a} \otimes \mbf{b}$ with $[\mbf{H}]_{ij} = [\mbf{a}]_i [\mbf{b}]_j$.
Operation of a second order tensor $\mbf{H}$ on a vector $\mbf{a}$ is given by $[\mbf{H} \mbf{a}]_i = [\mbf{H}]_{ij} [\mbf{a}]_j$.
Scalar product of two tensors $\mbf{H}$ and $\mbf{G}$ is denoted as $\mbf{H} \cdot \mbf{G} = [\mbf{H}]_{ij}[\mbf{G}]_{ij} $.
A list of key variables used throughout this manuscript is presented in Table \ref{table: notation}.
}

\begin{table}[h]
   \caption{Notation used in this manuscript.}
   \label{table: notation}
{\tabulinesep=1.2mm
   \begin{tabu} {c l | c l }
       \hline
  $\mbf{x}$ & Position vector (spatial) & $\mbf{X}$ & Position vector (referential) \\ 
  \hline 
  $\mbf{n}$ & Unit outward normal (spatial)&  $\mbf{n}_0$ & Unit outward normal (referential) \\
  \hline 
  $\phi$ & Electric scalar potential (spatial) & $\Phi$ & Electric scalar potential (referential) \\
  \hline
  $\potC$ & Electric vector potential (spatial) & $\potR$ & Electric vector potential (referential) \\
  \hline
  $\elecC$ & Electric field vector (spatial) & $\elecR$ & Electric field vector (referential) \\
  \hline
  $\disC$ & Electric displacement vector (spatial) & $\disR$ & Electric displacement vector (referential) \\
  \hline
  $\polC$ & Electric polarization vector (spatial) & $\polR$ & Electric polarization vector (referential) \\
  \hline
  $\strC$ & Cauchy stress tensor & $\strPK$ & First Piola--Kirchhoff stress tensor \\
  \hline
  $\jump{\{ \bullet \}}$ & Jump of a quantity across the boundary & $\{ \bullet \}_{,\mbf{G}} $ & Partial derivative with respect to $\mbf{G}$ \\
   & $\jump{\{ \bullet \}} = \{ \bullet \}^+ - \{ \bullet \}^- $ & & \\
   \hline
   \end{tabu}}
\end{table}

{For tensor calculus and variational method, we refer to \citep{Knowles1997, Itskov2018} and \citep{Gelfand2003, Giaquinta2010}, respectively, whereas the notation and definitions of physical entities in continuum mechanics typically follow \citep{Gurtin1981}.}
   
\section{Nonlinear electroelastostatics: some fundamental equations and entities} \label{sec: basic eqns}

\begin{figure}
\begin{center}
\includegraphics[width=0.7\linewidth]{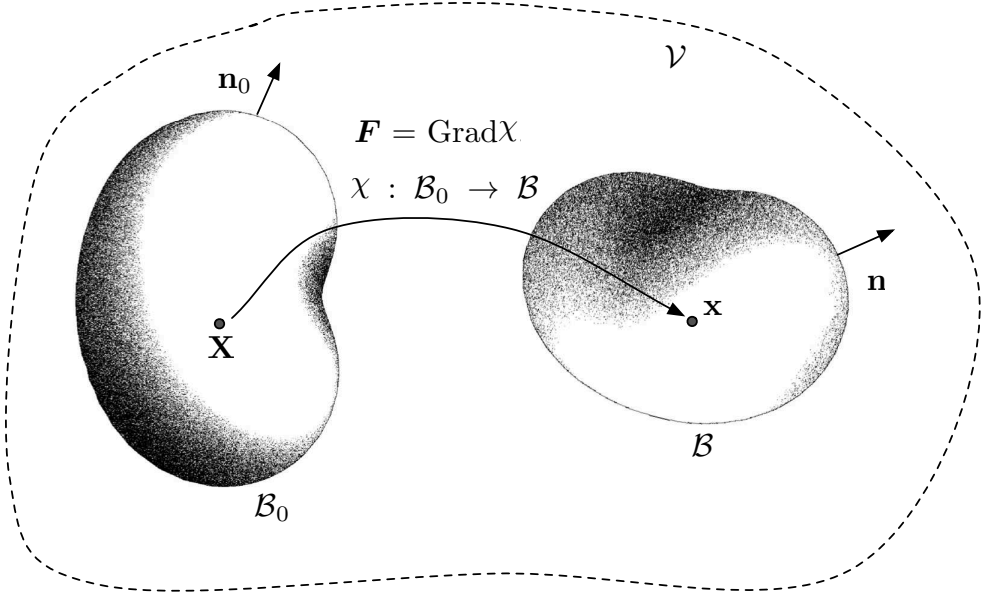}
\end{center}
\caption{A representation of the problem depicting the body in its reference and current configurations embedded in a volume $\mcal{V}$.}
\label{fig: problem cartoon}
\end{figure}

Consider a deformable body  {absent of free surface or volume charges} occupying a domain $\bod$ lying inside a region $\vol$ as schematically depicted in Figure \ref{fig: problem cartoon}. We denote the exterior of the body relative to $\vol$ by $$\bod' = \vol \setminus (\bod \cup \bbod).$$ 
We assume that the body occupies a domain $\bod_0$ in its reference configuration.
The points in domains $\bod_0$ and $\bod$ corresponding to the same material point of the body are naturally mapped into each other by the deformation function $$\rchi: \bod_0 \to \bod.$$
In order to make sense of the Lagrangian description of fields in current region $\vol$, but outside the body, in a meaningful manner, we also define an extension of the deformation function $\rchi$ to the part of domain outside the body such that sufficient continuity requirements are maintained. Thus, by an abuse of notation, we assume an extension of mapping $\rchi$ on a larger region, also denoted by $$\rchi: \vol_0 \to \vol,$$
where $\vol_0$ is the referential region corresponding to $\vol$.
This concept of a {\em fictitious} deformation function was {initially} used by \cite{Toupin1956} (and possibly others) for similar problems.
Following standard notation {in continuum mechanics}, we define the deformation gradient as $\Dgrad = \text{Grad} \rchi$.
The extension of $\rchi$ to $\vol_0$ allows us to define the exterior of the body in the reference configuration; this is denoted by $$\bod_0' = \vol_0 \setminus (\bod_0 \cup \bbod_0).$$

The electric field vector, electric displacement vector, and the electric polarization vector are denoted in the reference configuration as $(\elecR, \disR, \polR)$, respectively and in the current configuration as $(\elecC, \disC, \polC)$.
These three vector fields are related by the well known constitutive relation
\begin{equation}
\disC = \varepsilon_0 \elecC + \polC. \label{eqn: constitutive d e p}
\end{equation}
Further, the vector fields $(\elecC, \disC, \polC)$ must satisfy the Maxwell's equations
\begin{equation}
\text{div}\, \disC = 0, \quad \curl \elecC = \mbf{0}, \quad \quad \forall \, \mbf{x} \, \in \bod \cup \bod'. \label{eq: maxwell 1}
\end{equation}
The divergence-free and curl-free conditions \eqref{eq: maxwell 1} lead to the existence of electric potential (vector) field $\potC$ and electric potential (scalar) field $\phi$ on $\bod \cup \bod'$; the respective expressions of $\disC$ and $\elecC$ are given by
\begin{equation}
\disC = \curl \potC, \quad \quad \elecC = - \grad \phi.
\end{equation}
By using the Lagrangian counterparts of $\disC$ and $\elecC$, defined by
\begin{equation}
\disR = J \Dgrad^{-1} \disC, \quad \quad \elecR = \Dgrad^\top \elecC, \label{eqn: Dd, Ee relations}
\end{equation}
we rewrite the Maxwell's equations \eqref{eq: maxwell 1} in the reference configuration as
\begin{equation}
\Div \disR = 0, \quad \Curl \elecR = \mbf{0}, \quad \quad \forall \, \mbf{X} \, \in \bod_0 \cup \bod'_0. \label{eqn: maxwell lagrangian}
\end{equation}
Based on the referential equations \eqref{eqn: maxwell lagrangian}, we also define the suitable Lagrangian counterparts of the electric vector potential and electric scalar potential on $\bod_0 \cup \bod'_0$ as
\begin{equation}
\disR = \Curl \potR, \quad \elecR = - \Grad \Phi. \label{eqn: potentials introduction}
\end{equation}
{It can be shown using tensor algebra and calculus that}
\begin{equation}
\potR(\mbf{X}) = \Dgrad^\top(\mbf{X}) \potC(\mbf{x}) \bigg|_{\mbf{x} = \rchi (\mbf{X})}, \quad \quad \Phi(\mbf{X}) = \phi(\mbf{x}) \bigg|_{\mbf{x} = \rchi (\mbf{X})}, 
\end{equation}
for all $\mbf{X} \, \in \bod_0 \cup \bod'_0$.
Upon substituting the transformations \eqref{eqn: Dd, Ee relations} into the constitutive relation \eqref{eqn: constitutive d e p}, we get
\begin{equation}
J^{-1} \CGright \disR = \varepsilon_0 \elecR + \polR, \label{eqn: constitutive D E P}
\end{equation}
where $\polR$ denotes the Lagrangian electric polarization vector field that relates to the current electric polarization vector field by
\begin{equation}
\polR(\mbf{X}) = \Dgrad^\top(\mbf{X}) \polC(\mbf{x}) \bigg|_{\mbf{x}= \rchi (\mbf{X})},
\end{equation}
for all $\mbf{X} \, \in \bod_0 \cup \bod'_0$ (as $\polC$ is zero in $\bod'$, we also get vanishing $\polR$ in $\bod'_0$).

\section{Variational formulation based on electric displacement $\disR$}
\label{sec: D formulation}

Using the fact that $\disR$ is found in terms of $\potR$ by \eqref{eqn: potentials introduction}${}_1$, i.e., $\disR = \Curl \potR$,
the total potential energy of the system, i.e., the body and its exterior, is written as a functional depending on the deformation $\rchi$ and $\potR$ as \citep{Dorfmann2014b}
\begin{align}
E[\rchi, \potR] &= \int\limits_{\bod_0} \Omega (\Dgrad, \disR) dv_0 + \frac{1}{2 \varepsilon_0} \int\limits_{\bod'_0} J^{-1} [\Dgrad \disR] \cdot [\Dgrad \disR] dv_0 + \int\limits_{\partial \vol} \left[ \elecC_a \wedge \potC \right]\cdot \mbf{n} ds \nonumber \\
& - \int\limits_{\bod_0} \wt{\mbf{f}}^e \cdot \rchi \, dv_0 - \int\limits_{\bbod_0} \wt{\mbf{t}}^e \cdot \rchi \, ds_0, 
\label{eqn: E 1}
\end{align}
where
$\Omega$ is the (scalar)  {total} stored energy density per unit volume, 
$\elecC_a$ is the externally applied electric (vector) field whose tangential component is prescribed on $\partial \mcal{V}$, $\wt{\mbf{f}}^e$ is the body force (vector) field per unit volume while $\wt{\mbf{t}}^e$ is the applied traction (vector) field  {due to dead loads} at the boundary of the body in its current configuration. 
In \eqref{eqn: E 1} the integration is defined on the reference configuration and the current fields are mapped to the reference by using the mapping $\rchi$ as placement.
%
%
The exception is the third term in equation \eqref{eqn: E 1}, which is written in terms of the current region $\vol$.
However, it assumed that the boundary (typically, {\em infinitally} far away) is fixed (i.e., it does not change in space between reference and current description), so that the third term in equation \eqref{eqn: E 1} is also rewritten in the reference configuration simply as
\begin{equation}
\int\limits_{\partial \vol_0} \left[ \elecR_a \wedge \potR \right] \cdot \mbf{n}_0 ds_0.
\label{eqn: E 1 aux}
\end{equation}
Notice that $\mbf{n}_0$ and $\mbf{n}$ are used to denote the outward unit normals for the region $\vol_0$ and $\vol$ as well.
%
%

\subsection{Equilibrium: first variation}
\label{sec: A first var}
In order to describe $\rchi$ and $\potR$ when the body is in a state of equilibrium, the first variation of the energy functional should vanish, that is, using the functional \eqref{eqn: E 1}
\begin{equation}
\delta E\equiv \delta E[\rchi, \potR; (\delta \rchi, \delta \potR)] = 0. \label{eqn: first variation condition}
\end{equation}
An expansion of the functional $E$ up to the first order, owing to a variation of its arguments $\rchi$ and $\potR$, is given by
\begin{align}
E [\rchi + \delta \rchi, \potR + \delta \potR] &= \int\limits_{\bod_0} \Omega (\Dgrad +\delta \Dgrad, \disR + \delta \disR) dv_0 \nonumber \\
&+ \frac{1}{2 \varepsilon_0} \int\limits_{\bod'_0} [J + \delta J]^{-1} \left[ [\Dgrad + \delta \Dgrad] \left[\disR + \delta \disR \right] \right] \cdot \left[ [ \Dgrad + \delta \Dgrad] \left[\disR + \delta \disR \right] \right] dv_0 \nonumber \\
&+ \int\limits_{\partial \vol_0} \left[ \elecR_a \wedge [\potR + \delta \potR] \right] \cdot \mbf{n}_0 ds_0 - \int\limits_{\bod_0} \wt{\mbf{f}}^e \cdot [ \rchi+ \delta \rchi ] \, dv_0 - \int\limits_{\bbod_0} \wt{\mbf{t}}^e \cdot [ \rchi+ \delta \rchi ] \, ds_0.
\end{align}
Taking advantage of the referential description, noting that
\begin{equation}
\delta \mbf{D} = \text{Curl} \delta \potR,
\end{equation}
while using expressions for first order variations stated in the form of Appendix \ref{appendix: variations},
we simplify further the expression of $E [\rchi + \delta \rchi, \potR + \delta \potR] $ stated above. Thus, it is found that the first variation \eqref{eqn: first variation condition}
of $E$ is given by
\begin{align}
\delta E &= E [\rchi + \delta \rchi, \potR + \delta \potR] - E [\rchi, \potR ] \nonumber\\
&= \int\limits_{\bod_0} \left[ \Omega_{,\Dgrad} \cdot \delta \Dgrad + \Omega_{, \disR} \cdot \text{Curl} \delta \potR \right] dv_0 \nonumber \\
& + \frac{1}{2 \varepsilon_0} \int\limits_{\bod'_0} \bigg[ - J^{-1} \left[ \Dgrad^{-\top} \cdot \delta \Dgrad \right] [\Dgrad \disR] \cdot [\Dgrad \disR] + 2 J^{-1} [[\Dgrad \disR ] \otimes \disR] \cdot \delta \Dgrad + 2 [\CGright \disR] \cdot \text{Curl} \delta \potR \bigg] dv_0 \nonumber \\
& + \int\limits_{\partial \vol_0} \left[ \mbf{n}_0 \wedge \elecR_a \right] \cdot \delta \potR ds_0 - \int\limits_{\bod_0} \wt{\mbf{f}}^e \cdot \delta \rchi \, dv_0 - \int\limits_{\bbod_0} \wt{\mbf{t}}^e \cdot \delta \rchi \, ds_0.
\end{align}
Using an elementary identity for vector fields $\mbf{u}$ and $\mbf{v}$, namely,
\begin{equation}
 { \mbf{v} \cdot \Curl \mbf{u}   = \Div [\mbf{u} \wedge \mbf{v}] + [\Curl \mbf{v}] \cdot \mbf{u}, } \label{eqn: curl identity}
\end{equation}
we expand the above expression for $ \delta E$ as
\begin{align}
\delta E &= \int\limits_{\bod_0} \left[ \Omega_{,\Dgrad} \cdot \delta \Dgrad + [ { \Curl }\Omega_{, \disR} ] \cdot \delta \potR \right] dv_0 + \int\limits_{\bbod_0^{-}} \mbf{n}_0 \cdot \left[ \Omega_{,\disR} \wedge \delta \potR \right] ds_0 \nonumber \\
& - \frac{1}{\varepsilon_0} \int\limits_{\bbod_0^+} \mbf{n}_0 \cdot \left[ \CGright \disR \wedge \delta \mbf{A} \right] ds_0 + \frac{1}{2 \epsilon_0} \int\limits_{\bod'_0} \bigg[ - J^{-1} \left[ \Dgrad^{-\top} \cdot \delta \Dgrad \right] [\Dgrad \disR] \cdot [\Dgrad \disR] + 2 J^{-1} [[\Dgrad \disR ] \otimes \disR] \cdot \delta \Dgrad \nonumber \\
& + [\text{Curl}( \CGright \disR)] \cdot \delta \potR \bigg] \, dv_0 + \int\limits_{\partial \vol_0} \left[ \mbf{n}_0 \wedge \left[ \elecR_a - \frac{1}{\varepsilon_0} \CGright \disR \right] \right] \cdot \delta \potR ds_0 - \int\limits_{\bod_0} \wt{\mbf{f}}^e \cdot \delta \rchi \, dv_0 - \int\limits_{\bbod_0} \wt{\mbf{t}}^e \cdot \delta \rchi \, ds_0.
\end{align}
Inspection of above leads to consideration of the definition of a tensor field given by
\begin{align}
\mbf{P}_m 
& = \frac{1}{ \varepsilon_0 J} \bigg[ [\Dgrad \disR ] \otimes [\Dgrad \disR ] - \frac{1}{2} [\Dgrad \disR] \cdot [\Dgrad \disR] \mbf{I} \bigg] \Dgrad^{-\top}, \label{eqn: maxwell strPK}
\end{align}
{where $\mbf{I}$ is a two-point identity tensor.}
Using the definition \eqref{eqn: maxwell strPK}, we rewrite the first variation $\delta E$ of the total potential as
\begin{align}
\delta E &= \int\limits_{\bod_0} \left[ - \left[ \text{Div} \left( \Omega_{,\Dgrad} \right) + \wt{\mbf{f}}^e \right] \cdot \delta \rchi + [ { \Curl} \Omega_{, \disR} ] \cdot \delta \potR \right] dv_0 \nonumber \\
& + \int\limits_{\bbod_0} \Bigg[ \left[ \left[ \Omega_{,\Dgrad} \bigg|_- - \mbf{P}_m \bigg|_+ \right] \mbf{n}_0 - \wt{\mbf{t}}^e \right] \cdot \delta \rchi + \left[ \mbf{n}_0 \wedge \left[ \Omega_{, \disR} \Big|_- - \frac{1}{\epsilon_0} \CGright \disR \Big|_+ \right] \right] \cdot \delta \potR \Bigg] \, ds_0 \nonumber \\
& + \int\limits_{\bod'_0} \bigg[- \text{Div} \, \mbf{P}_m \cdot \delta \rchi 
+ \frac{1}{2 \epsilon_0} [ { \Curl } \CGright \disR] \cdot \delta \potR \bigg] \, dv_0 \nonumber \\ 
&+ \int\limits_{\partial \vol_0} \bigg[ \mbf{P}_m \mbf{n}_0 \cdot \delta \rchi + \left[ \mbf{n}_0 \wedge \left[ \elecR_a - \frac{1}{\varepsilon_0} \CGright \disR \right] \right] \cdot \delta \potR \bigg] ds_0. 
\label{eqn: first variation condition1}
\end{align}
Following the traditional definition, the  {total} (first Piola--Kirchhoff) stress $\mbf{P}$ in the body is
\begin{equation}
\mbf{P} = \Omega_{,\Dgrad}, \quad \text{in} \quad \bod_0,
\label{eqn: piola strPK}
\end{equation}
and the (Maxwell) stress outside the body is given by \eqref{eqn: maxwell strPK}, i.e.,
\begin{equation}
\mbf{P} = \mbf{P}_m, \quad \text{in} \quad \bod_0'.
\end{equation}

Upon applying the condition \eqref{eqn: first variation condition} to the first variation \eqref{eqn: first variation condition1} calculated above, the coefficients of arbitrary variations $\delta \rchi$ and $\delta \potR$ should vanish for $\delta E$ to vanish.
Vanishing of the coefficients of $\delta \rchi$ results in the following equations
\begin{subequations}
\begin{align}
\text{Div} \, \mbf{P} + \wt{\mbf{f}}^e = \mbf{0}, \quad &\text{in} \quad \bod_0, \label{eqn: gov 1 D formulation} \\
\text{Div} \, \mbf{P} = \mbf{0}, \quad &\text{in} \quad \bod_0',\\
\jump{\mbf{P}} \mbf{n}_0 + \wt{\mbf{t}}^e= \mbf{0}, \quad &\text{on} \quad \bbod_0, \\
\mbf{P} \mbf{n}_0 = \mbf{0}, \quad &\text{on} \quad \partial \vol_0.
\end{align}
\label{eqn: Euler Eq D}
\end{subequations}


We define the electric field $\elecR$ in the body as
\begin{equation}
\elecR = \Omega_{, \disR} = \frac{1}{\varepsilon_0} \left[ J^{-1} \CGright \disR - \polR \right], \quad \text{in} \quad \bod_0,
\label{eqn: E = omegaD}
\end{equation}
and outside the body as
\begin{equation}
\elecR = \frac{1}{\varepsilon_0 }J^{-1} \CGright \disR, \quad \text{in} \quad \bod_0',
\end{equation}
because the electric polarization $\polR$ vanishes in $\bod_0'$ and use has been made of the constitutive relation \eqref{eqn: constitutive D E P}.
Since the body $\bod_0$ and normal to the boundary $\mbf{n}_0$ can be chosen arbitrarily, we get the following relations from the vanishing of the coefficients of $\delta \pot$ 
\begin{subequations}
\begin{align}
\text{Curl}( \elecR ) = \mbf{0}, \quad &\text{in} \quad \bod_0 \cup \bod_0', \\
\mbf{n}_0 \wedge \jump{\elecR} = \mbf{0}, \quad &\text{on} \quad \bbod_0, \\
\mbf{n}_0 \wedge \left[ \elecR_a - \elecR \right] = \mbf{0}, \quad &\text{on} \quad \partial \vol_0. \label{eqn: gov last D formulation}
\end{align}
\label{eqn: Euler Eq 2 D}
\end{subequations}

\begin{sidenote}
The Cauchy stress $\strC$ in the body is related to the first Piola--Kirchhoff \eqref{eqn: piola strPK} by the Piola transform as
\begin{equation}
\strC \, \text{cof}(\Dgrad) = \mbf{P}. \label{eqn: Piola transform stress}
\end{equation}
Upon using the relation \eqref{eqn: Dd, Ee relations}$_1$ and the tensor field stated as \eqref{eqn: maxwell strPK}, the counterpart $\strC_m$ of the Cauchy stress $\strC$ in $\bod'$ (vacuum) is given by the expression
\begin{equation}
\strC = \strC_m = { \frac{1}{ \varepsilon_0} \left[ \disC \otimes \disC - \frac{1}{2} [\disC \cdot \disC] \bsym{i} \right] \quad \text{in} \quad \bod', } \label{eqn: Maxwell Cauchy stress expression in vacuum}
\end{equation}
{where $\bsym{i}$ is the spatial identity tensor.}
\end{sidenote}
\begin{sidenote}
We note that in this formulation based on the electric displacement vector, we have apriori assumed that the equation \eqref{eqn: maxwell lagrangian}$_1$ is satisfied by $\disR$ and have recovered the equation \eqref{eqn: maxwell lagrangian}$_2$ for the electric field $\elecR$ as the Euler-Lagrange equation for the variational (energy minimisation) problem.
This procedure implies the constitutive assumption $\elecR = \Omega_{,\disR}$; the same constitutive assumption has been also independently derived using the second law of thermodynamics  \citep{Dorfmann2005a} .
\label{note1}
\end{sidenote}

\subsection{Critical point: second variation}
\label{second variation 1}


For the analysis of critical point $(\rchi, \potR)$, we need to find the functions $\Delta \rchi$ and $\Delta \potR$ such that the bilinear functional defined below vanishes at the critical point, that is
\begin{equation}
\delta^2 E\equiv \delta^2 E[\rchi, \potR; (\delta \rchi, \delta \potR), (\Delta \rchi, \Delta \potR)]
= 0. 
\label{eqn: second variation condition}
\end{equation}
Upon using the expressions derived in Appendix \ref{appendix: taylor for two variables}, the bilinear functional associated with the second variation \eqref{eqn: second variation condition}
of $E$ is expanded into the form
\begin{align}
\delta^2 E
& = \int\limits_{\bod_0} \Bigg[ \bigg[ \Omega_{, \Dgrad \Dgrad} \Delta \Dgrad + \frac{1}{2} \Omega_{, \Dgrad \disR} \Delta \disR + \frac{1}{2} \wt\Omega_{ \Dgrad \disR} \Delta \disR \bigg] \cdot \delta \Dgrad \nonumber \\
& + \bigg[ \Omega_{, \disR \disR} \Delta \disR + \frac{1}{2} \Omega_{, \disR \Dgrad} \Delta \Dgrad + \frac{1}{2} \wt\Omega_{ \disR \Dgrad} \Delta \Dgrad \bigg] \cdot \delta \disR \Bigg] dv_0 \nonumber \\
& + \frac{1}{2 \varepsilon_0} \int\limits_{\bod_0'} J^{-1} \Bigg[ [\Dgrad \disR] \cdot [\Dgrad \disR] \bigg[ \big[ \Dgrad^{-\top} \cdot \Delta \Dgrad \big] \big[ \Dgrad^{-\top} \cdot \delta \Dgrad \big] + \Dgrad^{-\top} [\Delta \Dgrad]^\top \Dgrad^{-\top} \cdot \delta \Dgrad \nonumber \\
& - 2 \bigg[ \Big[ \Delta \Dgrad \disR \Big] \cdot \big[ \Dgrad \disR \big] + \Big[ \Dgrad \Delta \disR \Big] \cdot \big[ \Dgrad \disR \big] \bigg] \Dgrad^{-\top} \cdot \delta \Dgrad \nonumber \\
& - 2 \bigg[ \Big[ \delta \Dgrad \disR \Big] \cdot \big[ \Dgrad \disR \big] + \Big[ \Dgrad \delta \disR \Big] \cdot \big[ \Dgrad \disR \big] \bigg] \Dgrad^{-\top} \cdot \Delta \Dgrad \nonumber \\
& + 2 \big[ \delta \Dgrad \Delta \disR + \Delta \Dgrad \delta \disR \big] \cdot \big[ \Dgrad \disR \big] + 2 \delta \Dgrad \disR \cdot \Dgrad \Delta \disR + 2 \Delta \Dgrad \disR \cdot \Dgrad \delta \disR \nonumber \\
& + 2 \big[ \Delta \Dgrad \disR \big] \cdot \big[ \delta \Dgrad \disR \big] + 2 \big[ \Dgrad \Delta \disR \big] \cdot \big[ \Dgrad \delta \disR \big] \Bigg] dv_0. \label{eqn: expanded d2E D formulation}
\end{align}
In the expression stated above we have defined the third order tensors $\wt{\Omega}_{\Dgrad \disR}$ and $\wt{\Omega}_{\disR \Dgrad }$ according to the following property
\begin{equation}
\left[ \wt{\Omega}_{\Dgrad \disR} \mbf{u} \right] \cdot \mbf{U} = \left[ \Omega_{, \disR \Dgrad} \mbf{U} \right] \cdot \mbf{u}, \quad \left[ \wt{\Omega}_{ \disR \Dgrad} \mbf{U} \right] \cdot \mbf{u} = \left[ \Omega_{, \Dgrad \disR} \mbf{u} \right] \cdot \mbf{U},
\end{equation}
which holds for arbitrary $\mbf{u}$ and $\mbf{U}$, while $\mbf{u}$ is a vector and $\mbf{U}$ is a second order tensor.
Using the expression \eqref{eqn: expanded d2E D formulation} of $\delta^2 E$, in the region $\bod_0'$ the terms containing $\delta \disR$ can be written in the form $\mbf{v}_0 \cdot \delta \disR$ where
\begin{align}
\mbf{v}_0 = \frac{1}{ \varepsilon_0 J} \bigg[ - \big[ \Dgrad^{-\top} \cdot \Delta \Dgrad \big] \Dgrad^\top \Dgrad \disR + [\Delta \Dgrad]^\top \Dgrad \disR + \Dgrad^\top \Delta \Dgrad \disR + \Dgrad^\top \Dgrad \Delta \disR \bigg].
\end{align}
Since equation \eqref{eqn: constitutive D E P} gives $\elecR = J^{-1} \varepsilon_0^{-1} \CGright \disR$ in $\bod_0'$, it is easy to see that 
\begin{align}
\mbf{v}_0 = \Delta \elecR.
\end{align}
Also, in the expression \eqref{eqn: expanded d2E D formulation} of $\delta^2 E$, in the region $\bod_0'$ the terms containing $\delta \Dgrad$ can be written in the form $\mbf{T} \cdot \delta \Dgrad$ where
\begin{align}
\mbf{T} & = \frac{1}{2 \varepsilon_0 J} \Bigg[ [\Dgrad \disR] \cdot [\Dgrad \disR] \bigg[ \big[ \Dgrad^{-\top} \cdot \Delta \Dgrad \big] \Dgrad^{-\top} + \Dgrad^{-\top} [\Delta \Dgrad]^\top \Dgrad^{-\top} \bigg] \nonumber \\
& - 2 \bigg[ \Big[ \Delta \Dgrad \disR \Big] \cdot \big[ \Dgrad \disR \big] + \Big[ \Dgrad \Delta \disR \Big] \cdot \big[ \Dgrad \disR \big] \bigg] \Dgrad^{-\top} - 2 \big[ \Dgrad^{-\top} \cdot \Delta \Dgrad \big] [\Dgrad \disR] \otimes \disR \nonumber \\
& + 2 [\Dgrad \disR] \otimes \Delta \disR + 2 [\Dgrad \Delta \disR] \otimes \disR + 2 [\Delta \Dgrad \disR] \otimes \disR \Bigg].
\label{eqn: delta Pm expression 1}
\end{align}
By expanding the expression stated in equation \eqref{eqn: maxwell strPK}, to first order perturbation, it is seen that 
\begin{align}
\mbf{T} = \Delta \strPK_m.
\end{align}
With the details provided in Appendix \ref{appendix: details 1} based on repeated application of the triple product identity involving the curl operator \eqref{eqn: curl identity} and the divergence theorem, while observing that the variations $\delta \rchi$ and $\delta \pot$ are arbitrary, the equation $\delta^2 E=0$ \eqref{eqn: expanded d2E D formulation} finally leads to the following partial differential equations
\begin{subequations}
\begin{align}
\text{Div} \left( \Omega_{, \Dgrad \Dgrad} \Delta \Dgrad + \frac{1}{2} \left[ \Omega_{, \Dgrad \disR} + \wt{\Omega}_{ \Dgrad \disR} \right] \Delta \disR \right) = 0 \quad &\text{in} \quad \bod_0, \\
\text{Curl} \left( \Omega_{, \disR \disR} \Delta \disR + \frac{1}{2} \left[ \Omega_{, \disR \Dgrad} + \wt{\Omega}_{\disR \Dgrad } \right] \Delta \Dgrad \right) = 0 \quad &\text{in} \quad \bod_0, \\
\bigg[ \left[ \Omega_{, \Dgrad \Dgrad} \Delta \Dgrad + \frac{1}{2} \left[ \Omega_{, \Dgrad \disR} + \wt{\Omega}_{ \Dgrad \disR} \right] \Delta \disR \right] \bigg|_- - \mbf{T} \bigg|_+ \bigg] \mbf{n}_0 = 0 \quad &\text{on} \quad \bbod_0,\\
\bigg[ \Omega_{, \disR \disR} \Delta \disR + \frac{1}{2} \left[ \Omega_{, \disR \Dgrad} + \wt{\Omega}_{\disR \Dgrad } \right] \Delta \Dgrad \bigg|_- - \mbf{v}_0 \bigg|_+ \bigg] \wedge \mbf{n}_0 = 0 \quad &\text{on} \quad \bbod_0,\\
\text{Div} (\mbf{T}) = 0 \quad &\text{in} \quad \bod_0',\\
\text{Curl}(\mbf{v}_0) = 0 \quad &\text{in} \quad \bod_0',\\
\mbf{T} \mbf{n}_0 = 0 \quad &\text{on} \quad \partial \vol_0,\\
\mbf{v}_0 \wedge \mbf{n}_0 = 0 \quad &\text{on} \quad \partial \vol_0.
\end{align}
\label{eqn: second variation based PDE}
\end{subequations}
The set of equations \eqref{eqn: second variation based PDE} need to be solved for the non-trivial unknown functions $(\Delta \rchi, \Delta \potR)$ describing the onset of bifurcation.

\begin{sidenote}
Note that since we have proved $\mbf{T} = \Delta \strPK_m$ and $\mbf{v}_0 = \Delta \elecR$, it also follows that the above set of equations for the variations $\Delta \disR$ and $\Delta \Dgrad$  {in $\bod_0'$} can be alternatively obtained by  {perturbing} the corresponding equations of equilibrium \eqref{eqn: gov 1 D formulation}--\eqref{eqn: gov last D formulation}.
 {However, perturbation of the governing equations in $\bod_0$ do not result in the above equations due to presence of the $ \frac{1}{2} \left[ \Omega_{, \Dgrad \disR} + \wt{\Omega}_{ \Dgrad \disR} \right]$ and $\frac{1}{2} \left[ \Omega_{, \disR \Dgrad} + \wt{\Omega}_{\disR \Dgrad } \right]$ terms. 
This general argument  can be relaxed in cases when the energy density function $\Omega$ is assumed to be sufficiently continuous as has been considered, for example, by \cite{Dorfmann2010a, Dorfmann2014a}. }
\label{perturbeqlb1}
\end{sidenote}

\begin{sidenote}
We note in passing that the analysis presented above can be extended to include the special case of incompressibility in a straightforward manner.
The assumption of incompressibility is equivalent to the constraint $ J - 1=0 $ in $\bod_0 $. Hence, we consider a modified energy function which includes one more term 
\begin{equation}
g(\Dgrad) = p [J-1],
\end{equation}
in the integrand of total energy denisty in bulk. In this modified energy function, the scalar field $p$ is recognized as the Lagrange multiplier associated with the incompressibility constraint.
Due to a variation $\delta \rchi$, we get the following Taylor's expansion for $g$
\begin{align}
g(\Dgrad + \delta \Dgrad) &= p \big[ J - 1 + \delta J + \delta^2 J \big] \nonumber \\
&= p \Bigg[ J - 1 + J \Dgrad^{-\top} \cdot \delta \Dgrad + \frac{1}{2} J \bigg[ \big[ \Dgrad^{-\top} \cdot \delta \Dgrad \big] \big[ \Dgrad^{-\top} \cdot \delta \Dgrad \big] - \Dgrad^{-\top} \big[ \delta \Dgrad]^\top \Dgrad^{-\top} \cdot \delta \Dgrad \bigg] \Bigg].
\end{align}

Substituting the above in the first variation of total potential energy functional and setting $J=1$, we get the following updated constitutive equation for the first Piola--Kirchhoff stress
\begin{equation}
\strPK = \Omega_{,\Dgrad} + p \Dgrad^{-\top}.
\end{equation}
\label{incompress}
\end{sidenote}

%
%
%
%

\section{Variational formulation based on electric field $\bbbm{E}$}
\label{sec: E formulation}


Noting that $\elecR = - \Grad \Phi$, the total potential energy of the system is written as \citep{Dorfmann2014b}
\begin{align}
E[\rchi, \Phi] &= \int\limits_{\bod_0} \wt{\Omega} (\Dgrad, \elecR) dv_0 - \frac{1}{2 }\varepsilon_0 \int\limits_{\bod'_0} J \left[ \Dgrad^{-\top} \elecR \right] \cdot \left[ \Dgrad^{-\top} \elecR \right] dv_0 - \int\limits_{\partial \vol} \phi \disC_a \cdot \mbf{n} ds \nonumber \\
& - \int\limits_{\bod_0} \wt{\mbf{f}}^e \cdot \rchi \, dv_0 - \int\limits_{\bbod_0} \wt{\mbf{t}}^e \cdot \rchi \, ds_0, 
\label{eqn: chi phi energy functional 1}
\end{align}
where
$\wt\Omega$ is the stored energy density per unit volume that depends on the deformation gradient $\Dgrad$ and the referential electric displacement vector $\elecR$.
$\disC_a$ is the externally applied electric displacement whose normal component is prescribed on $\partial \mcal{V}$, $\wt{\mbf{f}}^e$ is the body force per unit volume while $\wt{\mbf{t}}^e$ is the applied traction at the boundary.
The third term in equation \eqref{eqn: chi phi energy functional 1} is in the current configuration but the same argument as that preceding \eqref{eqn: E 1 aux} allows it to be rewritten in the reference configuration as
%
%
%
%
\begin{equation}
-\int\limits_{\partial \vol_0} \Phi \disR_a \cdot \mbf{n}_0 \, ds_0.
\label{eqn: chi phi energy functional 1 aux}
\end{equation}

\subsection{Equilibrium: first variation}
At state of equilibrium, $\rchi$ and $\Phi$ are such that the first variation of the energy functional vanishes satisfying an analogue of equation \eqref{eqn: first variation condition}, i.e.,
\begin{equation}
\delta E\equiv \delta E[\rchi, \Phi; (\delta \rchi, \delta \Phi)] = 0. \label{eqn: first variation condition E}
\end{equation}
The variation of the functional $E$ up to the first order in $(\delta \rchi, \delta \Phi)$ is given by
%
\begin{align}
\delta E &= E [\rchi + \delta \rchi, \Phi + \delta \Phi] - E [\rchi, \Phi ] = \int\limits_{\bod_0} \left[ \wt\Omega_{,\Dgrad} \cdot \delta \Dgrad - \wt\Omega_{, \elecR} \cdot \Grad \delta \Phi \right] dv_0 \nonumber \\ 
& - \frac{1}{2 }\varepsilon_0 \int\limits_{\bod'_0} \bigg[ J \Dgrad^{-\top} \cdot \delta \Dgrad \left[ \Dgrad^{-\top} \elecR \right] \cdot \left[ \Dgrad^{-\top} \elecR \right] -2 J \left[ \Dgrad^{-\top} [\delta \Dgrad]^{\top} \Dgrad^{-\top} \elecR \right] \cdot \left[ \Dgrad^{-\top} \elecR \right] \nonumber \\
& + 2 J \left[ \Dgrad^{-\top} \elecR \right] \cdot \left[ \Dgrad^{-\top} \delta \elecR \right] \bigg] dv_0 - \int\limits_{\partial \vol_0} \delta \Phi\, \disR_a \cdot \mbf{n}_0 \, ds_0 - \int\limits_{\bod_0} \wt{\mbf{f}}^e \cdot \delta \rchi \, dv_0 - \int\limits_{\bbod_0} \wt{\mbf{t}}^e \cdot \delta \rchi \, ds_0. 
\label{eqn: E first var 1}
\end{align}
We define the first Piola--Kirchhoff stress $\strPK$ and electric displacement $\disR$ in the body as
\begin{equation}
\strPK = \wt\Omega_{,\Dgrad}, \quad \quad \disR = - \wt\Omega_{, \elecR} \quad \quad \text{in} \quad \bod_0,
\label{eq: PK D}
\end{equation}
the (Maxwell) stress $\strPK_m$ outside the body as used earlier in equation \eqref{eqn: maxwell strPK}
and recall the relation $J^{-1} \Dgrad \disR = \varepsilon_0 \Dgrad^{-\top} \elecR$ in vacuum from equation \eqref{eqn: constitutive D E P}. 
Using the above relations \eqref{eq: PK D}, we rewrite the first variation \eqref{eqn: E first var 1} as
\begin{align}
\delta E& = \int\limits_{\bod_0} \bigg[ \Div \left( \strPK^{\top} \delta \rchi \right) - \left[ \Div \strPK + \wt{\mbf{f}}^e \right] \cdot \delta \rchi + \Div \left( \delta \Phi \, \disR \right) - \delta \Phi \, \Div \disR \bigg] dv_0 \nonumber \\
& + \int\limits_{\bod'_0} \bigg[ \Div \left( \strPK^{\top}_m \delta \rchi \right) - \left[ \Div \strPK_m \right] \cdot \delta \rchi + \Div \left( \delta \Phi \, \disR \right) - \delta \Phi \, \Div \disR \bigg] dv_0 \nonumber \\
& - \int\limits_{\partial \vol_0} \delta \Phi\, \disR_a \cdot \mbf{n}_0 \, ds_0 - \int\limits_{\bbod_0} \wt{\mbf{t}}^e \cdot \delta \rchi \, ds_0.
\label{eqn: E first var 12}
\end{align}
After an application of divergence theorem 
to \eqref{eqn: E first var 12}, we get
\begin{align}
\delta E &= \int\limits_{\bod_0} \bigg[ - \left[ \Div (\strPK) + \wt{\mbf{f}}^e \right] \cdot \delta \rchi - \delta \Phi \, \Div \disR \bigg] dv_0 \nonumber \\
& + \int\limits_{\bbod_0} \Bigg[ \bigg[ \bigg[ \strPK\big|_- - \strPK_m \big|_+ \bigg] \mbf{n}_0 - \wt{\mbf{t}}^e \bigg] \cdot \delta \rchi + \delta \Phi \bigg[ \disR \big|_- - \disR \big|_+ \bigg] \cdot \mbf{n}_0 \Bigg] ds_0 \nonumber \\
& + \int\limits_{\bod_0'} \Bigg[ - \left[ \Div \strPK_m \right] \cdot \delta \rchi - \delta \Phi \, \Div \disR \Bigg] dv_0 \nonumber \\
& + \int\limits_{\partial \vol_0} \bigg[ \strPK_m \mbf{n}_0 \cdot \delta \rchi + \delta \Phi \big[ \disR - \disR_a \big] \cdot \mbf{n}_0 \bigg] dv_0.
\label{eqn: E first var 13}
\end{align}
Since the two variations $\delta \rchi$ and $\delta \Phi$ are arbitrary, their coefficients in each of the integrals must vanish.
Accordingly, using the coefficient of $\delta \rchi$ in \eqref{eqn: E first var 13}, we get the equations
\begin{subequations}
\begin{align}
\text{Div} \, \mbf{P} + \wt{\mbf{f}}^e = \mbf{0}, \quad &\text{in} \quad \bod_0, \label{eqn: gov 1 E formulation}\\
\text{Div} \, \mbf{P} = \mbf{0}, \quad &\text{in} \quad \bod_0',\\
\jump{\mbf{P}} \mbf{n}_0 + \wt{\mbf{t}}^e= \mbf{0}, \quad &\text{on} \quad \bbod_0, \\
\mbf{P} \mbf{n}_0 = \mbf{0}, \quad &\text{on} \quad \partial \vol_0,
\end{align}
\label{eqn: Euler based E}
\end{subequations}
while the coefficient of $\delta \Phi$ in \eqref{eqn: E first var 13} leads to the equations
\begin{subequations}
\begin{align}
\Div \disR = 0, \quad &\text{in} \quad \bod_0,\\
\Div \disR = 0, \quad &\text{in} \quad \bod_0',\\
\jump{\disR} \cdot \mbf{n}_0 = 0, \quad &\text{on} \quad \bbod_0, \\
\jump{\disR} \cdot \mbf{n}_0 = 0, \quad &\text{on} \quad \partial \vol_0, \label{eqn: gov last E formulation}
\end{align}
\label{eqn: Euler 2 based E}
\end{subequations}

\begin{sidenote}
Parallel to the remark \ref{note1} at the end of Section \ref{sec: A first var},
we note that in this formulation based on the electric field (equivalently, the electric scalar potential), we have apriori assumed the equation \eqref{eqn: maxwell lagrangian}$_2$ that $\elecR$ should satisfy and have recovered the equation \eqref{eqn: maxwell lagrangian}$_1$ for the electric displacement $\disR$ as an Euler-Lagrange equation of this minimisation problem.
This procedure too implies the constitutive assumption $\disR = - \wt\Omega_{,\elecR}$ while it has been also independently derived earlier based on the second law of thermodynamics \citep{Saxena2014a}.
\label{note2}
\end{sidenote}

\begin{sidenote}
{
 We also note here that the two variational formulations based on $\disR$ and $\elecR$   can be related by applying a Legendre-type transform on the energy functions $\Omega$ and $\wt\Omega$ \citep{Dorfmann2005a}
\begin{equation}
 \Omega(\Dgrad, \disR) = \wt\Omega(\Dgrad, \elecR) + \disR \cdot \elecR. \label{eqn: legendre type transform}
\end{equation}
The above relations result in the electric constitutive relations \eqref{eqn: E = omegaD} and \eqref{eq: PK D}.
However, since $\disR$ and $\elecR$ are not dual variables (a third electric variable $\polR$ is also present), a proper Legendre transform to link $\Omega$ and $\wt\Omega$ is not readily available.
As such, the relation \eqref{eqn: legendre type transform} leads to different convexity properties for $\Omega$ and $\wt\Omega$ in general.
}
\end{sidenote}

\subsection{Critical point: second variation}

For the analysis of critical point $(\rchi, \Phi)$, we need to find $\Delta \rchi$ and $\Delta \Phi$ such that certain bilinear functional based on the second variation vanishes at the critical point, that is
\begin{equation}
\delta^2 E\equiv \delta^2 E[\rchi, \Phi; (\delta \rchi, \delta \Phi), (\Delta \rchi, \Delta \Phi)]
= 0. 
\label{eqn: second variation condition E}
\end{equation}
From the variational formulation based on the electric field $\bbbm{E}$ \eqref{eqn: chi phi energy functional 1}, using the expansions described in Appendix \ref{appendix: taylor for two variables}, we get the expanded expression for $ \delta^2 E$ as follows
\begin{align}
\delta^2 E &= \int\limits_{\bod_0} \Bigg[ \Div \left( \bigg[ \wt\Omega_{, \Dgrad \Dgrad} \Delta \Dgrad 
+ \frac{1}{2} \wt\Omega_{, \Dgrad \elecR} \Delta \elecR 
+ \frac{1}{2} \wh\Omega_{ \Dgrad \elecR } \Delta \elecR \bigg] ^\top \delta \rchi \right) \nonumber \\
& - \Div \left( \wt\Omega_{, \Dgrad \Dgrad} \Delta \Dgrad 
+ \frac{1}{2} \wt\Omega_{, \Dgrad \elecR} \Delta \elecR 
+ \frac{1}{2} \wh\Omega_{ \Dgrad \elecR } \Delta \elecR \right) \cdot \delta \rchi \nonumber \\
& - \Div \left( \bigg[ \frac{1}{2} \wh\Omega_{\elecR \Dgrad} \Delta \Dgrad
+ \frac{1}{2} \wt\Omega_{, \elecR \Dgrad} \Delta \Dgrad 
+ \wt\Omega_{, \elecR \elecR} \Delta \elecR
\bigg] \delta \Phi \right) \nonumber \\
& + \Div \left( \frac{1}{2} \wh\Omega_{\elecR \Dgrad} \Delta \Dgrad
+ \frac{1}{2} \wt\Omega_{, \elecR \Dgrad} \Delta \Dgrad 
+ \wt\Omega_{, \elecR \elecR} \Delta \elecR \right) \delta \Phi \Bigg] dv_0 \nonumber \\
& + \int\limits_{\bod_0'} \bigg[ \Div \left( \wt{\mbf{T}}^\top \delta \rchi \right) - \Div \wt{\mbf{T}} \cdot \delta \rchi + \Div \left( \wt{\mbf{v}}_0 \delta \Phi \right) - \Div \wt{\mbf{v}}_0 \, \delta \Phi \bigg] dv_0
\label{eqn: second variation condition E 1}, 
\end{align}
where we have introduced the tensor $\wt{\mbf{T}}$ and the vector $\wt{\mbf{v}}_0$ as
\begin{align}
\wt{\mbf{T}} = & {J}{\varepsilon_0} \Bigg[ \Dgrad^{-\top} [\Delta \Dgrad]^\top \Dgrad^{-\top} \elecR \otimes \Dgrad^{-1} \Dgrad^{-\top} \elecR + \Dgrad^{-\top} \elecR \otimes \Dgrad^{-1} \Delta \Dgrad \Dgrad^{-1} \Dgrad^{-\top} \elecR \nonumber \\
&- \Dgrad^{-\top} \Delta \elecR \otimes \Dgrad^{-1} \Dgrad^{-\top} \elecR - \Dgrad^{-\top} \elecR \otimes \Dgrad^{-1} \Dgrad^{-\top} \Delta \elecR \nonumber \\
& + \Dgrad^{-\top} \elecR \otimes \Dgrad^{-1} \Dgrad^{-\top} [\Delta \Dgrad]^\top \Dgrad^{-\top} \elecR - \big[ \Dgrad^{-\top} \cdot \Delta \Dgrad \big] \Dgrad^{-\top} \elecR \otimes \Dgrad^{-1} \Dgrad^{-\top} \elecR
\nonumber \\
& + \bigg[ - \big[ \Dgrad^{-\top} \left[ \Delta \Dgrad \right]^{\top} \Dgrad^{-\top} \elecR \big] \cdot \big[ \Dgrad^{-\top} \elecR \big] 
+ \big[ \Dgrad^{-\top} \elecR \big] \cdot \Dgrad^{-\top} \big[ \Delta \elecR \big] \bigg]\Dgrad^{-\top} \nonumber \\
& - \frac{1}{2} \big[ \Dgrad^{-\top} \elecR \big] \cdot \big[ \Dgrad^{-\top} \elecR \big] \bigg[ \big[ \Dgrad^{-\top} \cdot \Delta \Dgrad \big] \Dgrad^{-\top} - \Dgrad^{-\top} [\Delta \Dgrad]^\top \Dgrad^{-\top} \bigg ], \\
\wt{\mbf{v}}_0 = & {J}{\varepsilon_0} \bigg[ \Dgrad^{-1} \Delta \Dgrad \Dgrad^{-1} \Dgrad^{-\top} + \Dgrad^{-1} \Dgrad^{-\top} \big[ \Delta \Dgrad \big]^\top \Dgrad^{-\top} - \left[ \Dgrad^{-\top} \cdot \Delta \Dgrad \right] \Dgrad^{-1} \Dgrad^{-\top} \bigg] \elecR \nonumber \\
& - J \varepsilon_0 \Dgrad^{-1} \Dgrad^{-\top} \Delta \elecR,
\end{align}
while we have also utilized the definitions of two third order tensors $\wh{\Omega}_{ \Dgrad \elecR}$ and $\wh{\Omega}_{ \elecR \Dgrad }$, according to the relations
\begin{equation}
\left[ \wh{\Omega}_{\Dgrad \elecR} \mbf{u} \right] \cdot \mbf{U} = \left[  { \wt \Omega_{, \elecR \Dgrad}} \mbf{U} \right] \cdot \mbf{u}, \quad \left[ \wh{\Omega}_{ \elecR \Dgrad} \mbf{U} \right] \cdot \mbf{u} = \left[  { \wt\Omega_{, \Dgrad \elecR} } \mbf{u} \right] \cdot \mbf{U},
\end{equation}
where $\mbf{u}$ and $\mbf{U}$ are arbitrary vector and arbitrary second order tensor, respectively.

An application of divergence theorem to \eqref{eqn: second variation condition E 1} gives
\begin{align}
\delta^2 E &= \int\limits_{\bod_0} \Bigg[ - \Div \left( \wt\Omega_{, \Dgrad \Dgrad} \Delta \Dgrad 
+ \frac{1}{2} \wt\Omega_{, \Dgrad \elecR} \Delta \elecR 
+ \frac{1}{2} \wh\Omega_{ \Dgrad \elecR } \Delta \elecR \right) \cdot \delta \rchi \nonumber \\
& + \Div \left( \frac{1}{2} \wh\Omega_{\elecR \Dgrad} \Delta \Dgrad
+ \frac{1}{2} \wt\Omega_{, \elecR \Dgrad} \Delta \Dgrad 
+ \wt\Omega_{, \elecR \elecR} \Delta \elecR \right) \delta \Phi \Bigg] dv_0 \nonumber \\
& + \int\limits_{\bbod_0} \Bigg[ \bigg[ \wt\Omega_{, \Dgrad \Dgrad} \Delta \Dgrad 
+ \frac{1}{2} \wt\Omega_{, \Dgrad \elecR} \Delta \elecR 
+ \frac{1}{2} \wh\Omega_{ \Dgrad \elecR } \Delta \elecR \bigg] \bigg|_- - \wt{\mbf{T}} \bigg|_+ \Bigg] \mbf{n}_0 \cdot \delta \rchi ds_0 \nonumber \\
& - \int\limits_{\bbod_0} \Bigg[ \bigg[ \frac{1}{2} \wh\Omega_{\elecR \Dgrad} \Delta \Dgrad
+ \frac{1}{2} \wt\Omega_{, \elecR \Dgrad} \Delta \Dgrad 
+ \wt\Omega_{, \elecR \elecR} \Delta \elecR \bigg] \bigg|_- - \wt{\mbf{v}}_0 \bigg|_+ \Bigg] \cdot \mbf{n}_0 \delta \Phi ds_0 \nonumber \\
& + \int\limits_{\bod_0'} \bigg[ - \Div \wt{\mbf{T}} \cdot \delta \rchi - \Div \wt{\mbf{v}}_0 \, \delta \Phi \bigg] dv_0 + \int\limits_{\partial \mcal{V}_0} \bigg[ \wt{\mbf{T}} \mbf{n}_0 \cdot \delta \rchi + \wt{\mbf{v}}_0 \cdot \mbf{n}_0 \delta \Phi \bigg] ds_0.
\end{align}
Since the variations $\delta \rchi$ and $\delta \Phi$ are arbitrary, we arrive at the following equations for the unknown functions $(\Delta \rchi, \Delta \Phi)$
\begin{subequations}
\begin{align}
\Div \left( \wt\Omega_{, \Dgrad \Dgrad} \Delta \Dgrad 
+ \frac{1}{2} \wt\Omega_{, \Dgrad \elecR} \Delta \elecR 
+ \frac{1}{2} \wh\Omega_{ \Dgrad \elecR } \Delta \elecR \right) & = \mbf{0} \quad \text{in} \quad \bod_0,\\
\Div \left( \frac{1}{2} \wh\Omega_{\elecR \Dgrad} \Delta \Dgrad
+ \frac{1}{2} \wt\Omega_{, \elecR \Dgrad} \Delta \Dgrad 
+ \wt\Omega_{, \elecR \elecR} \Delta \elecR \right) & = 0 \quad \text{in} \quad \bod_0,\\
\Bigg[ \bigg[ \wt\Omega_{, \Dgrad \Dgrad} \Delta \Dgrad 
+ \frac{1}{2} \wt\Omega_{, \Dgrad \elecR} \Delta \elecR 
+ \frac{1}{2} \wh\Omega_{ \Dgrad \elecR } \Delta \elecR \bigg] \bigg|_- - \wt{\mbf{T}} \bigg|_+ \Bigg] \mbf{n}_0 & = \mbf{0} \quad \text{on} \quad \bbod_0,\\
\Bigg[ \bigg[ \frac{1}{2} \wh\Omega_{\elecR \Dgrad} \Delta \Dgrad
+ \frac{1}{2} \wt\Omega_{, \elecR \Dgrad} \Delta \Dgrad 
+ \wt\Omega_{, \elecR \elecR} \Delta \elecR \bigg] \bigg|_- - \wt{\mbf{v}}_0 \bigg|_+ \Bigg] \cdot \mbf{n}_0 & = 0 \quad \text{on} \quad \bbod_0,\\
\Div (\wt{\mbf{T}}) & = \mbf{0} \quad \text{in} \quad \bod_0',\\
\Div (\wt{\mbf{v}}_0) & = {0} \quad \text{in} \quad \bod_0',\\
\wt{\mbf{T}} \mbf{n}_0 & = \mbf{0} \quad \text{on} \quad \partial \mcal{V}_0,\\
\wt{\mbf{v}}_0 \cdot \mbf{n}_0 & = {0} \quad \text{on} \quad \partial \mcal{V}_0,
\end{align}
\label{eqn: Perturb based E}
\end{subequations}
describing the onset of bifurcation.

\begin{sidenote}
Note that a variation of the relation $ \disR = J \varepsilon_0 \CGright^{-1} \elecR $ from equation \eqref{eqn: constitutive D E P} gives
\begin{align}
\Delta \disR= \wt{\mbf{v}}_0,
\label{eqn E v0}
\end{align}
\begin{align}
\text{since }
\Delta \disR&= J \varepsilon_0 \bigg[ \Dgrad^{-1} \Dgrad^{-\top} \Delta \elecR - \Dgrad^{-1} \Delta \Dgrad \Dgrad^{-1} \Dgrad^{-\top} \elecR - \Dgrad^{-1} \Dgrad^{-\top} [\Delta \Dgrad]^\top \Dgrad^{- \top} \elecR \nonumber \\
&+ \big[ \Dgrad^{-\top} \cdot \Delta \Dgrad \big] \Dgrad^{-1} \Dgrad^{-\top} \bigg].
\end{align}
A variation of the Maxwell stress \eqref{eqn: maxwell strPK} (after writing it in terms of $\elecR$ using the relation \eqref{eqn: constitutive D E P}) gives
\begin{align}
\Delta \strPK_m& = \wt{\mbf{T}},
\label{eqn E T}
\end{align}
\begin{align}
\text{since }
\Delta \strPK_m & = J \varepsilon_0 \Bigg[ \Dgrad^{-\top} \Delta \elecR \otimes \Dgrad^{-1} \Dgrad^{-\top} \elecR + \Dgrad^{-\top} \elecR \otimes \Dgrad^{-1} \Dgrad^{-\top} \Delta \elecR \nonumber \\
& + \big[ \Dgrad^{-\top} \cdot \Delta \Dgrad \big] \Dgrad^{-\top} \elecR \otimes \Dgrad^{-1} \Dgrad^{-\top} \elecR - \Dgrad^{-\top} [\Delta \Dgrad]^{ -\top} \elecR \otimes \Dgrad^{- 1} \Dgrad^{-\top} \elecR \nonumber \\
& - \Dgrad^{-\top} \elecR \otimes \Dgrad^{-1} \Dgrad^{-\top} [\Delta \Dgrad]^\top \Dgrad^{-\top} \elecR - \Dgrad^{-\top} \elecR \otimes \Dgrad^{-1} [\Delta \Dgrad ] \Dgrad^{-1 } \Dgrad^{-\top} \elecR \nonumber \\
& + \frac{1}{2} [\Dgrad^{-\top} \elecR] \cdot [\Dgrad^{-\top} \elecR] \big[ \Dgrad^{-\top} [\Delta \Dgrad]^\top \Dgrad^{-\top} - [\Dgrad^{-\top} \cdot \Delta \Dgrad] \Dgrad^{-\top} \big] \nonumber \\
& + \bigg[ - \big[ \Dgrad^{-\top} \left[ \Delta \Dgrad \right]^{\top} \Dgrad^{-\top} \elecR \big] \cdot \big[ \Dgrad^{-\top} \elecR \big] 
+ \big[ \Dgrad^{-\top} \elecR \big] \cdot \Dgrad^{-\top} \big[ \Delta \elecR \big] \bigg] \Dgrad^{-\top} \Bigg].
\label{eqn: delta Pm expression 2}
\end{align}
Alternative to the statements $\wt{\mbf{v}}_0 = \Delta \elecR$ \eqref{eqn E v0} and $\wt{\mbf{T}} = \Delta \strPK_m$ \eqref{eqn E T}, it can be also shown that the above set of  equations for the perturbations $\Delta \elecR$ and $\Delta \Dgrad$ can be obtained by linearising the equations of equilibrium \eqref{eqn: gov 1 E formulation}--\eqref{eqn: gov last E formulation}.
\label{perturbeqlb2}
\end{sidenote}

\section{Variational formulation based on electric polarization $\polR$}
\label{sec: P formulation}

Consider the body $\bod_0$ in its reference configuration in a space $\vol_0$.
Noting that $\elecR = -\Grad \Phi$,
the total potential energy of the system is given as \citep{Liu2014}
\begin{align}
E[\rchi, \polR] &= \int\limits_{\bod_0} \wh{\Omega} (\mbf{F}, \polR) \, dv_0 + \frac{\epsilon_0}{2} \int\limits_{\vol_0 } J \Big| \Dgrad^{-\top} \Grad { \Phi } \Big|^2 \, dv_0 \nonumber \\
& - \int\limits_{\bod_0} \wt{\mbf{f}}^e \cdot \rchi \, dv_0 - \int\limits_{\partial \bod_0}
\wt{\mbf{t}}^e \cdot \rchi \, ds_0 + \int\limits_{\partial \mcal{V}_0} \phi_b \mbf{n}_0 \cdot \disR \, ds_0, \label{eqn: potential energy functional for P formulation}
\end{align}
where
$\wh\Omega$ is the stored energy density per unit volume that depends on the deformation gradient $\Dgrad$ and the referential electric polarization vector $\polR$.
$\phi_b$ is the externally applied electric potential, 
$\wt{\mbf{f}}^e$ is the body force per unit volume while $\wt{\mbf{t}}^e$ is the applied traction at the boundary. 
{Unlike the previous two formulations, the energy in the region outside $\bod_0$ does not have a direct dependence on the independent variable $\polR$.
Thus taking first variation of this functional requires a different treatment than the procedure adopted in the previous sections and is presented below.}

\subsection{Equilibrium: first variation}
In order for a solution $\rchi$ and $\polR$ to be at equilibrium, the first variation of the energy functional should vanish satisfying equation \eqref{eqn: first variation condition}.
The variation of functional $E$ up to the first order is given by
\begin{align}
\delta E &= E[\rchi + \delta \rchi, \polR + \delta \polR] - E[\rchi, \polR] = \int\limits_{\bod_0} \Big[ \wh{\Omega}_{,\Dgrad} \cdot \delta \Dgrad + \wh{\Omega}_{,\polR} \cdot \delta \polR \Big] dv_0 \nonumber \\
& - \int\limits_{\bod_0} \wt{\mbf{f}}^e \cdot \delta \rchi \, dv_0 - \int\limits_{\bbod_0} \wt{\mbf{t}}^e \cdot \delta \rchi \, ds_0 + \int\limits_{\vol_0} \Bigg[ - \wh{\strPK}_m \cdot \delta \Dgrad - J \varepsilon_0 \big[ \CGright^{-1} \elecR \big] \cdot \Grad \delta \Phi \Bigg] dv_0 + \int\limits_{\partial \mcal{V}_0} \phi_b \mbf{n}_0 \cdot \delta \disR \, ds_0. 
\label{eqn: P form delta E 1}
\end{align}
where $\wh{\strPK}_m$ is the tensor defined below
\begin{equation}
\wh{\strPK}_m = \varepsilon_0 J \bigg[ - \frac{1}{2} \big[ \Dgrad^{-\top} \elecR \big] \cdot \big[ \Dgrad^{-\top} \elecR \big] \mbf{I} + \big[ \Dgrad^{-\top} \elecR \big] \otimes\big[ \Dgrad^{-\top} \elecR \big] \bigg] \Dgrad^{-\top}.
\label{eqn: max type tensor 1}
\end{equation}
Notice that outside the body, the electric polarization $\polR = \mbf{0}$, that gives $\wh{\strPK}_m = \strPK_m$, $\strPK_m$ being the Maxwell stress tensor defined in equation \eqref{eqn: maxwell strPK}.

We use the divergence theorem on the last term { of \eqref{eqn: P form delta E 1}} and use the condition from a variation of equation \eqref{eqn: maxwell lagrangian}$_1$ that $\Div(\delta \disR) = 0$ to get
\begin{equation}
\int\limits_{\partial \vol_0} \mbf{n}_0 \cdot \phi \delta \disR \, ds_0 = \int\limits_{\vol_0} \Div \left(\phi \, \delta \disR \right) \, dv_0 = \int\limits_{\vol_0} \Grad(\phi) \cdot \delta \disR \, dv_0 = - \int\limits_{\vol_0} \elecR \cdot \delta \disR \, dv_0. \label{eqn: last term P functional}
\end{equation}
%
%
{Using the constitutive relation \eqref{eqn: constitutive D E P}, an increment of electric displacement $\disR$ up to first order can be written as}
\begin{align}
\delta \disR 
& = \Big[ [\Dgrad^{-\top} \cdot \delta \Dgrad ] \mbf{I} - \CGright^{-1} [\delta \Dgrad]^\top \Dgrad - \Dgrad^{-1} [\delta \Dgrad] \Big] \disR 
- \varepsilon_0 J \CGright^{-1} \Grad \delta \Phi + J \CGright^{-1} \delta \polR. \label{eqn: Delta d}
\end{align}
{Upon substituting \eqref{eqn: last term P functional} and \eqref{eqn: Delta d} } in the last term of equation \eqref{eqn: P form delta E 1}, we get
\begin{align}
\delta E &= \int\limits_{\bod_0} \Big[ \wh{\Omega}_{,\Dgrad} \cdot \delta \Dgrad + \wh{\Omega}_{,\polR} \cdot \delta \polR - \wt{\mbf{f}}^e \cdot \delta \rchi \Big] dv_0 - \int\limits_{\bbod_0} \wt{\mbf{t}}^e \cdot \delta \rchi \, ds_0 \nonumber \\
& + \int\limits_{\vol_0} \bigg[ \Big[ \wt{\strPK}_m- \wh{\strPK}_m \Big] \cdot \delta \Dgrad - J \CGright^{-1} \elecR \cdot \delta \polR \bigg] dv_0, \label{eqn: delta E for P formulation 1}
\end{align}
where we have defined the tensor
\begin{align}
\wt{\strPK}_m &= \bigg[ - [\disR \cdot \elecR] \mbf{I} + [\Dgrad \disR] \otimes [\Dgrad^{-\top} \elecR] + [\Dgrad^{-\top} \elecR] \otimes [\Dgrad \disR] \bigg] \Dgrad^{-\top}, \\
&= 2 \wh{\strPK}_m + J \bigg[ - [\CGright^{-1} \polR \cdot \elecR] \mbf{I} + [\Dgrad^{-\top} \polR] \otimes [\Dgrad^{-\top} \elecR] + [\Dgrad^{-\top} \elecR] \otimes [\Dgrad^{-\top} \polR] \bigg] \Dgrad^{-\top}. 
\label{eqn: max type tensor 2}
\end{align}
In {the region $\bod_0'$}, $\polR = \mbf{0}$ which leads to $\wt{\strPK}_m = 2 \strPK_m$.

{Upon separating the integral over $\vol_0$ in \eqref{eqn: delta E for P formulation 1}  to two integrals on $\bod_0$ and $\bod_0'$, we obtain }
\begin{align}
\delta E &= \int\limits_{\bod_0} \Bigg[ \bigg[\wh{\Omega}_{\Dgrad} + \wt{\strPK}_m- \wh{\strPK}_m \bigg] \cdot \delta \Dgrad - \wt{\mbf{f}}^e \cdot \delta \rchi + \bigg[ \wh\Omega_{,\polR} - J \CGright^{-1} \elecR \bigg] \cdot \delta \polR \Bigg] dv_0 - \int\limits_{\bbod_0} \wt{\mbf{t}}^e \cdot \delta \rchi \, ds_0 \nonumber \\
& + \int\limits_{\bod_0'} \strPK_m \cdot \delta \Dgrad \, dv_0.
\end{align}
This is rewritten with the use of divergence theorem as
\begin{align}
\delta E &= \int\limits_{\bod_0} \bigg[ - \Big[ \text{Div} \left( \wh\Omega_{,\Dgrad} + \wt{\strPK}_m- \wh{\strPK}_m \right) + \wt{\mbf{f}}^e \Big] \cdot \delta \rchi + \bigg[ \wh\Omega_{,\polR} - J \CGright^{-1} \elecR \bigg] \cdot \delta \polR \Bigg] dv_0 \nonumber \\
& + \int\limits_{\bbod_0} \Bigg[ \bigg[ \Big[ \wh\Omega_{,\Dgrad} + \wt{\strPK}_m- \wh{\strPK}_m \Big] \bigg|_- - \strPK_m \bigg|_+ \bigg] \mbf{n}_0 - \wt{\mbf{t}}^e \Bigg] \cdot \delta \rchi \, ds_0 - \int\limits_{\bod_0'} \Div \strPK_m 
\cdot \delta \rchi \, dv_0 + \int\limits_{\partial \vol_0} \strPK_m \mbf{n}_0 \cdot \delta \rchi \, ds_0.
\end{align}
We define the first Piola--Kirchhoff stress in the body as
\begin{equation}
\strPK = \wh\Omega_{,\Dgrad} + \wt{\strPK}_m- \wh{\strPK}_m, \quad \text{in} \quad \bod_0, 
\label{eqn: first PK for P formulation 1}
\end{equation}
while we have the same Maxwell stress $\strPK = \strPK_m$ from equation \eqref{eqn: maxwell strPK} outside the body in $\bod_0'$ similar to what has been obtained in the other two formulations based on $\elecR$ and $\disR$.
Upon applying the condition \eqref{eqn: first variation condition} to the first variation calculated above, the coefficients of arbitrary variations $\delta \rchi$ and $\delta \polR$ should vanish for $\delta E$ to be zero.

Vanishing of the coefficients of $\delta \polR$ results in the following constitutive relation between $\elecR$ and $\polR$
\begin{equation}
\elecR = J^{-1} \CGright \wh{\Omega}_{, \polR} \quad \quad \text{in} \; \bod_0. \label{eqn: constitutive polarization electric field}
\end{equation}
{Upon substituting the above expression for $\elecR$ in equations \eqref{eqn: max type tensor 1}, \eqref{eqn: max type tensor 2}, and \eqref{eqn: first PK for P formulation 1}} 
the total first Piola--Kirchhoff stress can be rewritten in terms of the independent quantities $\Dgrad$ and $\polR$ as
\begin{align}
\strPK = \wh{\Omega}_{,\Dgrad} + \varepsilon_0 J^{-1} \Bigg[ - \frac{1}{2} \wh{\Omega}_{,\polR} \cdot \big[ \CGright \wh{\Omega}_{,\polR} \big] \mbf{I} + \wh{\Omega}_{,\polR} \otimes \big[ \CGright \wh{\Omega}_{,\polR} \big] \Bigg] \Dgrad^{-\top} \nonumber \\
 + \Bigg[ - \big[ \polR \cdot \wh{\Omega}_{,\polR} \big] \mbf{I} + \polR \otimes \wh{\Omega}_{,\polR} + \wh{\Omega}_{,\polR} \otimes \polR \Bigg] \Dgrad^{-\top}. \label{eqn: first PK in F and P terms}
\end{align}

Vanishing of the coefficients of $\delta \rchi$ results in the following equations
\begin{subequations}
\begin{align}
\text{Div} \, \mbf{P} + \wt{\mbf{f}}^e = \mbf{0}, \quad &\text{in} \quad \bod_0, \label{eqn: gov 1 P formulation} \\
\text{Div} \, \mbf{P} = \mbf{0}, \quad &\text{in} \quad \bod_0',\\
\jump{\mbf{P}} \mbf{n}_0 + \wt{\mbf{t}}^e= \mbf{0}, \quad &\text{on} \quad \bbod_0, \\
\mbf{P} \mbf{n}_0 = \mbf{0}, \quad &\text{on} \quad \partial \vol_0.
\end{align}
\label{eqn: Euler based P}
\end{subequations}

\begin{sidenote}
We note that in this formulation based on the electric polarization vector, we have to apriori use both the Maxwell's equations \eqref{eqn: maxwell lagrangian} to impose conditions on $\disR$ and $\elecR$ unlike the previous two formulations in which one condition was imposed and the other was derived.
Also unlike the previous two formulations, stress does not have a simple expression of being a derivative of the total energy density with respect to the deformation gradient tensor.
The procedure implies the constitutive relation \eqref{eqn: constitutive polarization electric field} between $\elecR$ and $\polR$.

\end{sidenote}

\subsection{Critical point: perturbation of equilibrium equation}

{For the analysis of critical point $(\rchi, \polR)$, the perturbations $\Delta \rchi$ and $\Delta \polR$ in the equilibrium state need to satisfy certain incremental equations and boundary conditions. These are derived from \eqref{eqn: Euler based P} and are stated below}
%
\begin{subequations}
\begin{align}
\text{Div} \, \Delta \mbf{P} = \mbf{0}, \quad &\text{in} \quad \bod_0, \label{eqn: incr gov 1 P formulation} \\
\text{Div} \, \Delta \mbf{P} = \mbf{0}, \quad &\text{in} \quad \bod_0',\\
\jump{\Delta \mbf{P}} \mbf{n}_0  = \mbf{0}, \quad &\text{on} \quad \bbod_0, \label{eqn: Perturb based P BC} \\
\Delta \mbf{P} \mbf{n}_0 = \mbf{0}, \quad &\text{on} \quad \partial \vol_0.
\end{align}
\label{eqn: Perturb based P}
\end{subequations}
We find a perturbation in the first Piola--Kirchhoff stress using equation \eqref{eqn: first PK in F and P terms} as
\begin{align}
\Delta \strPK& = \wh{\Omega}_{,\Dgrad \Dgrad} \Delta \Dgrad + \frac{1}{2} \Big[ \wh{\Omega}_{,\Dgrad \polR} + \wt{\Omega}_{\Dgrad \polR} \Big] \Delta \polR \nonumber \\
& - \varepsilon_0 J^{-1} \Big[ \Dgrad^{-\top} \cdot \Delta \Dgrad \Big] \Bigg[ - \frac{1}{2} \wh{\Omega}_{,\polR} \cdot \big[ \CGright \wh{\Omega}_{,\polR} \big] \mbf{I} + \wh{\Omega}_{,\polR} \otimes \big[ \CGright \wh{\Omega}_{,\polR} \big] \Bigg] \Dgrad^{-\top} \nonumber \\
& - \varepsilon_0 J^{-1} \Bigg[ - \frac{1}{2} \wh{\Omega}_{,\polR} \cdot \big[ \CGright \wh{\Omega}_{,\polR} \big] \mbf{I} + \wh{\Omega}_{,\polR} \otimes \big[ \CGright \wh{\Omega}_{,\polR} \big] \Bigg] \Dgrad^{-\top} [\Delta \Dgrad]^\top \Dgrad^{-\top} \nonumber \\
& + \varepsilon_0 J^{-1} \Bigg[ - \bigg[ \Dgrad \wh{\Omega}_{,\polR} \cdot \Big[ \Delta \Dgrad \wh{\Omega}_{,\polR} + \Dgrad \big[\wh{\Omega}_{,\polR \polR} \Delta \polR + \frac{1}{2} \wh{\Omega}_{,\polR \Dgrad} \Delta \Dgrad + \frac{1}{2} \wt{\Omega}_{\polR \Dgrad} \Delta \Dgrad \big] \Big] \bigg] \mbf{I} \nonumber \\
& + \big[\wh{\Omega}_{,\polR \polR} \Delta \polR + \frac{1}{2} \wh{\Omega}_{,\polR \Dgrad} \Delta \Dgrad + \frac{1}{2} \wt{\Omega}_{\polR \Dgrad} \Delta \Dgrad \big] \otimes \big[ \CGright \wh{\Omega}_{,\polR} \big] \nonumber \\
& + \wh{\Omega}_{,\polR} \otimes \bigg[ \CGright \big[\wh{\Omega}_{,\polR \polR} \Delta \polR + \frac{1}{2} \wh{\Omega}_{,\polR \Dgrad} \Delta \Dgrad + \frac{1}{2} \wt{\Omega}_{\polR \Dgrad} \Delta \Dgrad \big] + \Big[ [\Delta \Dgrad]^\top \Dgrad + \Dgrad^\top \Delta \Dgrad \Big] \wh{\Omega}_{,\polR} \bigg] \Bigg] \Dgrad^{-\top} \nonumber \\
& - \Bigg[ - \big[ \polR \cdot \wh{\Omega}_{,\polR} \big] \mbf{I} + \polR \otimes \wh{\Omega}_{,\polR} + \wh{\Omega}_{,\polR} \otimes \polR \Bigg] \Dgrad^{-\top} [\Delta \Dgrad]^\top \Dgrad^{-\top} \nonumber \\
& + \Bigg[ - \bigg[ \Delta \polR \cdot \wh{\Omega}_{, \polR} + \polR \cdot \Big[ \wh{\Omega}_{,\polR \polR} \Delta \polR + \frac{1}{2} \wh{\Omega}_{,\polR \Dgrad} \Delta \Dgrad + \frac{1}{2} \wt{\Omega}_{\polR \Dgrad} \Delta \Dgrad \Big] \bigg] \mbf{I} \nonumber \\
& + \Delta \polR \otimes \wh{\Omega}_{, \polR} + \polR \otimes \Big[ \wh{\Omega}_{,\polR \polR} \Delta \polR + \frac{1}{2} \wh{\Omega}_{,\polR \Dgrad} \Delta \Dgrad + \frac{1}{2} \wt{\Omega}_{\polR \Dgrad} \Delta \Dgrad \Big] \nonumber \\
& + \Big[ \wh{\Omega}_{,\polR \polR} \Delta \polR + \frac{1}{2} \wh{\Omega}_{,\polR \Dgrad} \Delta \Dgrad + \frac{1}{2} \wt{\Omega}_{\polR \Dgrad} \Delta \Dgrad \Big] \otimes \polR + \wh{\Omega}_{,\polR} \otimes \polR \Bigg] \Dgrad^{-\top},
\label{eqn: delta P for P formulation}
\end{align}
where we have defined two third order tensors $\wt{\Omega}_{ \Dgrad \polR}$ and $\wt{\Omega}_{ \polR \Dgrad }$ which have the following property
\begin{equation}
\left[ \wt{\Omega}_{\Dgrad \polR} \mbf{u} \right] \cdot \mbf{U} = \left[ \wh\Omega_{, \polR \Dgrad} \mbf{U} \right] \cdot \mbf{u}, \quad \left[ \wt{\Omega}_{ \polR \Dgrad} \mbf{U} \right] \cdot \mbf{u} = \left[ \wh\Omega_{, \Dgrad \polR} \mbf{u} \right] \cdot \mbf{U},
\end{equation}
$\mbf{u}$ being an arbitrary vector and $\mbf{U}$ being an arbitrary second order tensor.

{Perturbation in the Maxwell stress $\Delta \strPK_m$ in $\bod_0'$ in terms of $\Delta \Dgrad$ and $\Delta \elecR$ is given by Equation \eqref{eqn: delta Pm expression 2}.
The boundary condition \eqref{eqn: Perturb based P BC} connects $\Delta \strPK$  \eqref{eqn: delta P for P formulation} and $\Delta \strPK_m$ \eqref{eqn: delta Pm expression 2} through the constitutive relation \eqref{eqn: constitutive polarization electric field} for $\elecR$.

Notice that contrary to the previous two cases, in this formulation based on polarization we employ a direct perturbation based approach to derive equations for critical point instead of the second variation based analysis as the latter requires lengthy and convoluted manipulations.

}

\section{Concluding remarks}
The equations of nonlinear electroelastostatics have been analyzed using three different variational formulations with respect to the field variable for the electric effect, namely, the electric field $\elecR$, the electric displacement $\disR$, the electric polarization $\polR$. Although the first variation based Euler-Lagrange equation has been found to coincide with that documented in the published literature, it is the second variation based critical point analysis which brings a small surprise.
It is found that the second variation based partial differential equation satisfied by a perturbation near the critical point at bifurcation is not so straightforward.
After careful manipulations and simplifications of the several terms arising out of repeated application of divergence theorem and identities in vector calculus, we were able to obtain the relevant equations.
It is observed that there are certain terms which cannot be obtained by direct perturbation approach of the Euler--Lagrange equation.

\section*{Acknowledgements}
P.S. acknowledges the support of startup funds from the James Watt School of Engineering at the University of Glasgow.
B.L.S. acknowledges the support of SERB MATRICS grant MTR/2017/000013.

\appendix

\section{Variation of some relevant functions}
\label{appendix: variations}

In order to evaluate the first and second variations, we note the following relations on account of Taylor's expansion of relevant functions

Upon a variation in $\rchi \to \rchi + \delta \rchi$, we get
\begin{align}
\Dgrad (\rchi + \delta \rchi) = \text{Grad} \, \rchi + \text{Grad} \, (\delta \rchi) \quad \Rightarrow \quad \delta \Dgrad = \text{Grad} \, (\delta \rchi), \quad \delta^2 \Dgrad = \mbf{0}.
\end{align}
The Cauchy--Green deformation tensor will change as
\begin{align}
\CGright(\rchi + \delta \rchi) &= [\Dgrad + \delta \Dgrad]^{\top} [\Dgrad + \delta \Dgrad] = \Dgrad^{\top} \Dgrad + \Dgrad^{\top} \delta \Dgrad + [\delta \Dgrad]^{\top} \Dgrad + [\delta \Dgrad]^{\top} \delta \Dgrad, \nonumber \\
\Rightarrow \quad \delta \CGright &= \Dgrad^{\top} \delta \Dgrad + [\delta \Dgrad]^{\top} \Dgrad, \quad \delta^2 \CGright = [\delta \Dgrad]^{\top} \delta \Dgrad.
\end{align}

For the determinant $J$, we have
\begin{align}
J (\rchi + \delta \rchi ) &= J + \delta J + \delta^2 J + \dotsc \nonumber \\
&= \text{det} (\Dgrad + \delta \Dgrad) \nonumber \\
& = J + \text{cof} (\Dgrad) \cdot \delta \Dgrad + \Dgrad \cdot \text{cof}(\delta \Dgrad ) + \text{det} (\delta \Dgrad),
\end{align}
\begin{equation}
\Rightarrow \delta J = J \Dgrad^{-\top} \cdot \delta \Dgrad, \quad \quad \delta^2 J = \Dgrad \cdot \text{cof} (\delta \Dgrad).
\end{equation}
As $\delta \Dgrad = \text{Grad}( \delta \rchi)$, the second of the above expressions, $\delta^2 J$, is written in component form as
\begin{equation}
\delta^2 J = \frac{1}{2} \varepsilon_{imn} \varepsilon_{jpq} F_{ij} [\delta \rchi_{m,p} ] [\delta \rchi_{n,q}].
\end{equation}
{Here $\varepsilon_{ijk}$ is the third order permutation tensor}.
We present another more useful expression for second derivative of $J$,  simply obtained by differentiating the first derivative.
We write
\begin{equation}
\delta J = \pd{J}{\Dgrad} \cdot \delta \Dgrad, \quad \quad \Rightarrow \pd{J}{\Dgrad} = J \Dgrad^{-\top}.
\end{equation}
A directional derivative of the above expression gives
\begin{equation}
\frac{\partial}{\partial \Dgrad} \left( \pd{J}{\Dgrad} \right) \delta \Dgrad = J \big[ \Dgrad^{-\top} \cdot \delta \Dgrad \big] \Dgrad^{-\top} - J \Dgrad^{-\top} \big[ \delta \Dgrad]^\top \Dgrad^{-\top}.
\end{equation}
Thus, we have
\begin{equation}
\delta^2 J = \frac{1}{2} J \bigg[ \big[ \Dgrad^{-\top} \cdot \delta \Dgrad \big] \big[ \Dgrad^{-\top} \cdot \delta \Dgrad \big] - \Dgrad^{-\top} \big[ \delta \Dgrad]^\top \Dgrad^{-\top} \cdot \delta \Dgrad \bigg].
\label{eqn: modified expression for delta 2 J}
\end{equation}
Taylor's expansion for the inverse of determinant $J^{-1}$ is
\begin{equation}
J^{-1} (\rchi + \delta \rchi) = J_0 + J_1 + J_2 + \dotsc
\end{equation}
where
\begin{equation}
J_0 = J^{-1}, \quad J_1 = - J^{-1} \Dgrad^{-\top} \cdot \delta \Dgrad, \quad J_2 = - J^{-2} \Dgrad \cdot \text{cof} (\delta \Dgrad) + J^{-1} \left[ \Dgrad^{-\top} \cdot \delta \Dgrad \right]^2.
\end{equation}
Using the expression \eqref{eqn: modified expression for delta 2 J}, we rewrite $J_2$ as
\begin{equation}
J_2 = \frac{1}{2 J} \bigg[ \left[ \Dgrad^{-\top} \cdot \delta \Dgrad \right]^2 + \Dgrad^{-\top} [\delta \Dgrad]^\top \Dgrad^{-\top} \cdot \delta \Dgrad \bigg].
\end{equation}
For the inverse tensors, let
\begin{align}
[\Dgrad (\rchi + \delta \rchi) ]^{-1} = \Dgrad^{-1} + D_1 \Dgrad^{-1} + D_2 \Dgrad^{-1} + \dotsc
\end{align}
Comparing the terms of similar order in $\delta \Dgrad$ in
\begin{align}
[\Dgrad (\rchi + \delta \rchi) ]^{-1} [\Dgrad (\rchi + \delta \rchi) ] -\Dgrad^{-1} \Dgrad&=\mbf{I}-\mbf{I}=\mbf{0}\\
&= [ \Dgrad + \delta \Dgrad ] ^{-1} [ \Dgrad + \delta \Dgrad ] -\Dgrad^{-1} \Dgrad\\
&= \left[ \Dgrad^{-1} + D_1 \Dgrad^{-1} + D_2 \Dgrad^{-1} \right] [ \Dgrad + \delta \Dgrad ]-\Dgrad^{-1} \Dgrad,
\end{align}
we get
\begin{equation}
D_1 \Dgrad^{-1} = - \Dgrad^{-1} [ \delta \Dgrad ] \Dgrad^{-1}, \quad \quad D_2 \Dgrad^{-1} = \Dgrad^{-1} [ \delta \Dgrad ] \Dgrad^{-1} [ \delta \Dgrad ] \Dgrad^{-1}.
\end{equation}
For the inverse of the right Cauchy--Green deformation tensor $\CGright^{-1} = \Dgrad^{-1} \Dgrad^{-\top}$, let
\begin{align}
[\CGright (\rchi + \delta \rchi) ]^{-1} = \CGright^{-1} + D_1 \CGright^{-1} + D_2 \CGright^{-1} + \dotsc
\end{align}
Then, considering only the terms up to second order
\begin{equation}
\CGright^{-1} + D_1 \CGright^{-1} + D_2 \CGright^{-1} = \left[ \Dgrad^{-1} + D_1 \Dgrad^{-1} + D_2 \Dgrad^{-1} \right] \left[ \Dgrad^{-\top} + D_1 \Dgrad^{-\top} + D_2 \Dgrad^{-\top} \right],
\end{equation}
and comparing the terms of similar order in $\delta \Dgrad$, we get
\begin{align}
D_1 \CGright^{-1} &= - \CGright^{-1} [\delta \Dgrad]^{\top} \Dgrad^{-\top} - \Dgrad^{-1} [\delta \Dgrad] \CGright^{-1}, \\
D_2 \CGright^{-1} &= \CGright^{-1} [\delta \Dgrad]^{\top} \Dgrad^{-\top} [\delta \Dgrad]^{\top} \Dgrad^{-\top} + \Dgrad^{-1} [\delta \Dgrad] \CGright^{-1} [\delta \Dgrad]^{\top} \Dgrad^{-\top} + \Dgrad^{-1} [ \delta \Dgrad ] \Dgrad^{-1} [ \delta \Dgrad ] \CGright^{-1}.
\end{align}

\section{On functions with two separate types of variations}
\label{appendix: taylor for two variables}
Consider the energy density function $\Omega (\Dgrad, \elecR)$ and variations of the form $(\delta \Dgrad + \Delta \Dgrad)$ and $(\delta \elecR + \Delta \elecR)$.
Then
\begin{align}
\Omega (\Dgrad + \delta \Dgrad + \Delta \Dgrad, \elecR + \delta \elecR + \Delta \elecR) &= \Omega (\Dgrad, \elecR) + \Omega_{,\Dgrad} \cdot \left[ \delta \Dgrad + \Delta \Dgrad \right] + \Omega_{,\elecR} \cdot \left[ \delta \elecR + \Delta \elecR \right] \nonumber \\
& + \frac{1}{2} \bigg[\Omega_{,\Dgrad \Dgrad} \left[ \delta \Dgrad + \Delta \Dgrad \right] \bigg] \cdot \left[ \delta \Dgrad + \Delta \Dgrad \right] + \frac{1}{2} \bigg[\Omega_{,\Dgrad \elecR} \left[ \delta \elecR + \Delta \elecR \right] \bigg] \cdot \left[ \delta \Dgrad + \Delta \Dgrad \right] \nonumber \\
& + \frac{1}{2} \bigg[\Omega_{,\elecR \Dgrad} \left[ \delta \Dgrad + \Delta \Dgrad \right] \bigg] \cdot \left[ \delta \elecR + \Delta \elecR \right] + \frac{1}{2} \bigg[\Omega_{,\elecR \elecR} \left[ \delta \elecR + \Delta \elecR \right] \bigg] \cdot \left[ \delta \elecR + \Delta \elecR \right] .
\end{align}
Collecting only the second order terms and exploiting the major symmetries of $\Omega_{,\Dgrad \Dgrad}$ and $\Omega_{,\elecR \elecR}$, the second directional derivative $D_2\Omega$ is written as
\begin{align}
D_2 \Omega& = \frac{1}{2} \bigg[ \Omega_{,\Dgrad \Dgrad} \delta \Dgrad \bigg] \cdot \delta \Dgrad 
+ \bigg[ \Omega_{,\Dgrad \Dgrad} \Delta \Dgrad \bigg] \cdot \delta \Dgrad 
+ \frac{1}{2} \bigg[ \Omega_{,\Dgrad \Dgrad} \Delta \Dgrad \bigg] \cdot \Delta \Dgrad \nonumber \\
& + \frac{1}{2} \bigg[ \Omega_{,\Dgrad \elecR} \delta \elecR \bigg] \cdot \delta \Dgrad 
+ \frac{1}{2} \bigg[ \Omega_{,\Dgrad \elecR} \Delta \elecR \bigg] \cdot \delta \Dgrad
+ \frac{1}{2} \bigg[ \Omega_{,\Dgrad \elecR} \delta \elecR \bigg] \cdot \Delta \Dgrad
+ \frac{1}{2} \bigg[ \Omega_{,\Dgrad \elecR} \Delta \elecR \bigg] \cdot \Delta \Dgrad \nonumber \\
& + \frac{1}{2} \bigg[ \Omega_{,\elecR \Dgrad} \delta \Dgrad \bigg] \cdot \delta \elecR
+ \frac{1}{2} \bigg[ \Omega_{,\elecR \Dgrad} \Delta \Dgrad \bigg] \cdot \delta \elecR
+ \frac{1}{2} \bigg[ \Omega_{,\elecR \Dgrad} \delta \Dgrad \bigg] \cdot \Delta \elecR
+ \frac{1}{2} \bigg[ \Omega_{,\elecR \Dgrad} \Delta \Dgrad \bigg] \cdot \Delta \elecR \nonumber \\
& + \frac{1}{2} \bigg[ \Omega_{,\elecR \elecR} \delta \elecR \bigg] \cdot \delta \elecR 
+ \bigg[ \Omega_{,\elecR \elecR} \Delta \elecR \bigg] \cdot \delta \elecR
+ \frac{1}{2} \bigg[ \Omega_{,\elecR \elecR} \Delta \elecR \bigg] \cdot \Delta \elecR.
\end{align}
Now variations of the form $(\delta \Dgrad + \Delta \Dgrad)$ and $(\delta \elecR + \Delta \elecR)$ in equation \eqref{eqn: chi phi energy functional 1} gives for the integral over the region $\bod_0'$
\begin{align}
& - \frac{1}{2} \varepsilon_0 \int\limits_{\bod'_0} J \bigg[ 1 + \Dgrad^{- \top} \cdot \big[ \delta \Dgrad + \Delta \Dgrad \big] + \frac{1}{2} \big[ \Dgrad^{-\top} \cdot \big[ \delta \Dgrad + \Delta \Dgrad \big] \big] \big[ \Dgrad^{-\top} \cdot \big[ \delta \Dgrad + \Delta \Dgrad \big] \big] \nonumber \\
& - \frac{1}{2} \Dgrad^{-\top} \big[ \delta \Dgrad + \Delta \Dgrad \big]^\top \Dgrad^{-\top} \cdot \big[ \delta \Dgrad + \Delta \Dgrad \big] \bigg] \nonumber \\ 
&\Bigg[ \bigg[ \left[ \Dgrad + \delta \Dgrad + \Delta \Dgrad \right]^{-\top} \left[ \elecR + \delta \elecR + \Delta \elecR \right] \bigg] 
\cdot \bigg[ \left[ \Dgrad + \delta \Dgrad + \Delta \Dgrad \right]^{-\top} \left[ \elecR + \delta \elecR + \Delta \elecR \right] \bigg] \Bigg] dv_0.
\end{align}
Noting that
\begin{align}
\left[ \Dgrad + \delta \Dgrad + \Delta \Dgrad \right]^{-\top} = \Dgrad - \Dgrad^{-\top} \left[ \delta \Dgrad + \Delta \Dgrad \right]^{\top} \Dgrad^{-\top} 
+ \Dgrad^{-\top} \left[ \delta \Dgrad + \Delta \Dgrad \right]^{\top} \Dgrad^{-\top} \left[ \delta \Dgrad + \Delta \Dgrad \right]^{\top} \Dgrad^{-\top},
\end{align}
and collecting only the second order terms upon multiplication of the relevant terms, we get
\begin{align}
& - \frac{1}{2} \varepsilon_0 \int\limits_{\bod'_0} \Bigg[ 2 J \left[ \Dgrad^{-\top} \elecR \right] \cdot \bigg[ \Dgrad^{-\top} \left[ \delta \Dgrad + \Delta \Dgrad \right]^{\top} \Dgrad^{-\top} \left[ \delta \Dgrad + \Delta \Dgrad \right]^{\top} \Dgrad^{-\top} \elecR \bigg] \nonumber \\
& - 2 J \left[ \Dgrad^{-\top} \elecR \right] \cdot \bigg[ \Dgrad^{-\top} \left[ \delta \Dgrad + \Delta \Dgrad \right]^{\top} \Dgrad^{-\top} \left[ \delta \elecR + \Delta \elecR \right] \bigg] \nonumber \\
& - 2 J \bigg[ \Dgrad^{-\top} \left[ \delta \Dgrad + \Delta \Dgrad \right]^{\top} \Dgrad^{-\top} \elecR \bigg] \cdot \bigg[ \Dgrad^{-\top} \left[ \delta \elecR + \Delta \elecR \right] \bigg] \nonumber \\
& + J \bigg[ \Dgrad^{-\top} [\delta \elecR + \Delta \elecR] \bigg] \cdot \bigg[ \Dgrad^{-\top} [\delta \elecR + \Delta \elecR] \bigg] \nonumber \\
& + J \bigg[\Dgrad^{-\top} \left[ \delta \Dgrad + \Delta \Dgrad \right]^{\top} \Dgrad^{-\top} \elecR \bigg] \cdot \bigg[ \Dgrad^{-\top} \left[ \delta \Dgrad + \Delta \Dgrad \right]^{\top} \Dgrad^{-\top} \elecR \bigg] \nonumber \\
& + 2 J \Dgrad^{- \top} \cdot \big[ \delta \Dgrad + \Delta \Dgrad \big] \bigg[ - \big[ \Dgrad^{-\top} \left[ \delta \Dgrad + \Delta \Dgrad \right]^{\top} \Dgrad^{-\top} \elecR \big] \cdot \big[ \Dgrad^{-\top} \elecR \big] \nonumber \\
& + \big[ \Dgrad^{-\top} \elecR \big] \cdot \Dgrad^{-\top} \big[ \delta \elecR + \Delta \elecR \big] \bigg] \nonumber \\
& + \frac{1}{2} J\big[ \Dgrad^{-\top} \elecR \big] \cdot \big[ \Dgrad^{-\top} \elecR \big] \bigg[ \big[ \Dgrad^{-\top} \cdot \big[ \delta \Dgrad + \Delta \Dgrad \big] \big] \big[ \Dgrad^{-\top} \cdot \big[ \delta \Dgrad + \Delta \Dgrad \big] \big] \nonumber \\
& - \Dgrad^{-\top} \big[ \delta \Dgrad + \Delta \Dgrad \big]^\top \Dgrad^{-\top} \cdot \big[ \delta \Dgrad + \Delta \Dgrad \big] \bigg] \Bigg] dv_0.
\end{align}
Variations of the form $(\delta \Dgrad + \Delta \Dgrad)$ and $(\delta \disR + \Delta \disR)$ in equation \eqref{eqn: E 1} gives for the integral over the region $\bod_0'$
\begin{align}
& \frac{1}{2 \varepsilon_0} \int\limits_{\bod_0'} J^{-1} \Bigg[ 1 - \Dgrad^{-\top} \cdot \big[ \delta \Dgrad + \Delta \Dgrad \big] + \frac{1}{2} \Big[ \Dgrad^{-\top} \cdot \big[ \delta \Dgrad + \Delta \Dgrad \big] \Big] \Big[ \Dgrad^{-\top} \cdot \big[ \delta \Dgrad + \Delta \Dgrad \big] \Big] \nonumber \\
& + \frac{1}{2} \Dgrad^{-\top} \big[ \delta \Dgrad + \Delta \Dgrad \big]^\top \Dgrad^{-\top} \cdot \big[ \delta \Dgrad + \Delta \Dgrad \big] \Bigg] \nonumber \\
& \bigg[ \big[ \Dgrad + \delta \Dgrad + \Delta \Dgrad \big] \big[ \disR + \delta \disR + \Delta \disR \big] \cdot \big[ \Dgrad + \delta \Dgrad + \Delta \Dgrad \big] \big[ \disR + \delta \disR + \Delta \disR \big] \bigg] dv_0.
\end{align}
Upon collecting only the terms upto second order after multiplication, we get
\begin{align}
& \frac{1}{2 \varepsilon_0} \int\limits_{\bod_0'} \Bigg[ \frac{1}{2} J^{-1} [\Dgrad \disR] \cdot [\Dgrad \disR] \bigg[ \Big[ \Dgrad^{-\top} \cdot \big[ \delta \Dgrad + \Delta \Dgrad \big] \Big] \Big[ \Dgrad^{-\top} \cdot \big[ \delta \Dgrad + \Delta \Dgrad \big] \Big] \nonumber \\
& + \Dgrad^{-\top} \big[ \delta \Dgrad + \Delta \Dgrad \big]^\top \Dgrad^{-\top} \cdot \big[ \delta \Dgrad + \Delta \Dgrad \big] \bigg] \nonumber \\
& - 2 J^{-1} \Dgrad^{-\top} \cdot \big[ \delta \Dgrad + \Delta \Dgrad \big] \bigg[ \Big[ \big[ \delta \Dgrad + \Delta \Dgrad \big] \disR \Big] \cdot \big[ \Dgrad \disR \big] + \Big[ \Dgrad \big[ \delta \disR + \Delta \disR \big] \Big] \cdot \big[ \Dgrad \disR \big] \bigg] \nonumber \\
& + 2 J^{-1} \bigg[ \big[ \delta \Dgrad + \Delta \Dgrad \big] \big[ \delta \disR + \Delta \disR \big] \bigg] \cdot \big[ \Dgrad \disR \big] + 2 J^{-1} \Big[ \big[ \delta \Dgrad + \Delta \Dgrad \big] \disR \Big] \cdot \Big[ \Dgrad \big[ \delta \disR + \Delta \disR \big] \Big] \nonumber \\
& +J^{-1} \Big[ \big[ \delta \Dgrad + \Delta \Dgrad \big] \disR \Big] \cdot \Big[ \big[ \delta \Dgrad + \Delta \Dgrad \big] \disR \Big] + J^{-1} \Big[ \Dgrad \big[ \delta \disR + \Delta \disR \big] \Big] \cdot \Big[ \Dgrad \big[ \delta \disR + \Delta \disR \big] \Big] \Bigg] dv_0.
\end{align}


\section{Auxiliary details for calculations in \S\ref{second variation 1}}
\label{appendix: details 1}
Using the triple product identity involving the curl operator \eqref{eqn: curl identity}, we rewrite the equation $\delta^2 E=0$ \eqref{eqn: expanded d2E D formulation} as
\begin{align}
& \int\limits_{\bod_0} \bigg[ \text{Div} \left( [ \Omega_{, \Dgrad \Dgrad} \Delta \Dgrad + \frac{1}{2} \left[ \Omega_{, \Dgrad \disR} + \wt{\Omega}_{ \Dgrad \disR} \right] \Delta \disR ]^{\top} \delta \rchi \right) \nonumber \\
& - \left[ \text{Div} \left( \Omega_{, \Dgrad \Dgrad} \Delta \Dgrad + \frac{1}{2} \left[ \Omega_{, \Dgrad \disR} + \wt{\Omega}_{ \Dgrad \disR} \right] \Delta \disR \right) \right] \cdot \delta \rchi \nonumber \\
& + [\Omega_{, \disR \disR} \Delta \disR + \frac{1}{2} \left[ \Omega_{, \disR \Dgrad} + \wt{\Omega}_{\disR \Dgrad } \right] \Delta \Dgrad ] \cdot \delta \disR \bigg] dv_0 \nonumber \\
& + \int\limits_{\bod_0'} \bigg[ \text{Div} \left(\mbf{T}^{\top} \, \delta \rchi \right) - \left[ \text{Div} (\mbf{T})\right] \cdot \delta \rchi + \text{Div} \left( \delta \pot \wedge \mbf{v}_0 \right) + \text{Curl}(\mbf{v}_0) \cdot \delta \pot \bigg] dv_0 = 0. \label{eqn: second var Tv form}
\end{align}
By an application of the divergence theorem to \eqref{eqn: second var Tv form},
we get
\begin{align}
& \int\limits_{\bod_0} \left[ - \text{Div} \left( \Omega_{, \Dgrad \Dgrad} \Delta \Dgrad + \frac{1}{2} \left[ \Omega_{, \Dgrad \disR} + \wt{\Omega}_{ \Dgrad \disR} \right] \Delta \disR \right) \right] \cdot \delta \rchi \nonumber \\
& + [\Omega_{, \disR \disR} \Delta \disR + \frac{1}{2} \left[ \Omega_{, \disR \Dgrad} + \wt{\Omega}_{\disR \Dgrad } \right] \Delta \Dgrad ] \cdot \text{Curl} (\delta \pot) \bigg] dv_0 \nonumber \\
& +\int\limits_{\bbod_0^-} \left[ \Omega_{, \Dgrad \Dgrad} \Delta \Dgrad + \frac{1}{2} \left[ \Omega_{, \Dgrad \disR} + \wt{\Omega}_{ \Dgrad \disR} \right] \Delta \disR \right] \mbf{n}_0 \cdot \delta \rchi ds_0 \nonumber \\
& + \int\limits_{\bod_0'} \bigg[ - \text{Div} (\mbf{T}) \cdot \delta \rchi + \text{Div} \left( \delta \pot \wedge \mbf{v}_0 \right) + \text{Curl}(\mbf{v}_0) \cdot \delta \pot \bigg] dv_0 \nonumber \\
& + \int\limits_{\partial \vol_0} \mbf{T} \mbf{n}_0 \cdot \delta \rchi \, dv_0 - \int\limits_{\bbod_0^+} \mbf{T} \mbf{n}_0 \cdot \delta \rchi \, dv_0 = 0,
\end{align}
which can be simplified further by the identity \eqref{eqn: curl identity} so that
\begin{align}
& \int\limits_{\bod_0} \left[ - \text{Div} \left( \Omega_{, \Dgrad \Dgrad} \Delta \Dgrad + \frac{1}{2} \left[ \Omega_{, \Dgrad \disR} + \wt{\Omega}_{ \Dgrad \disR} \right] \Delta \disR \right) \right] \cdot \delta \rchi \nonumber \\
& + \text{Div} \left( \delta \pot \wedge [\Omega_{, \disR \disR} \Delta \disR + \frac{1}{2} \left[ \Omega_{, \disR \Dgrad} + \wt{\Omega}_{\disR \Dgrad } \right] \Delta \Dgrad ] \right) \nonumber \\
& + \text{Curl} \left( \Omega_{, \disR \disR} \Delta \disR + \frac{1}{2} \left[ \Omega_{, \disR \Dgrad} + \wt{\Omega}_{\disR \Dgrad } \right] \Delta \Dgrad \right) \cdot \delta \pot \bigg] dv_0 \nonumber \\
& +\int\limits_{\bbod_0^-} \left[ \Omega_{, \Dgrad \Dgrad} \Delta \Dgrad + \frac{1}{2} \left[ \Omega_{, \Dgrad \disR} + \wt{\Omega}_{ \Dgrad \disR} \right] \Delta \disR \right] \mbf{n}_0 \cdot \delta \rchi ds_0 \nonumber \\
& + \int\limits_{\bod_0'} \bigg[ - \text{Div} (\mbf{T}) \cdot \delta \rchi + \text{Div} \left( \delta \pot \wedge \mbf{v}_0 \right) + \text{Curl}(\mbf{v}_0) \cdot \delta \pot \bigg] dv_0 
\nonumber \\
& + \int\limits_{\partial \vol_0} \mbf{T} \mbf{n}_0 \cdot \delta \rchi \, dv_0 - \int\limits_{\bbod_0^+} \mbf{T} \mbf{n}_0 \cdot \delta \rchi \, dv_0 = 0.
\end{align}
Using the divergence theorem again
\begin{align}
& \int\limits_{\bod_0} \left[ - \text{Div} \left( \Omega_{, \Dgrad \Dgrad} \Delta \Dgrad + \frac{1}{2} \left[ \Omega_{, \Dgrad \disR} + \wt{\Omega}_{ \Dgrad \disR} \right] \Delta \disR \right) \right] \cdot \delta \rchi \nonumber \\
& + \text{Curl} \left( \Omega_{, \disR \disR} \Delta \disR + \frac{1}{2} \left[ \Omega_{, \disR \Dgrad} + \wt{\Omega}_{\disR \Dgrad } \right] \Delta \Dgrad \right) \cdot \delta \pot \bigg] dv_0 \nonumber \\
& + \int\limits_{\bod_0'} \bigg[ - \text{Div} (\mbf{T}) \cdot \delta \rchi + \text{Curl}(\mbf{v}_0) \cdot \delta \pot \bigg] dv_0 
\nonumber \\
& +\int\limits_{\bbod_0} \bigg[ \left[ \Omega_{, \Dgrad \Dgrad} \Delta \Dgrad + \frac{1}{2} \left[ \Omega_{, \Dgrad \disR} + \wt{\Omega}_{ \Dgrad \disR} \right] \Delta \disR \right] \bigg|_- - \mbf{T} \bigg|_+ \bigg] \mbf{n}_0 \cdot \delta \rchi ds_0 \nonumber \\
& + \int\limits_{\bbod_0} \bigg[ \Omega_{, \disR \disR} \Delta \disR + \frac{1}{2} \left[ \Omega_{, \disR \Dgrad} + \wt{\Omega}_{\disR \Dgrad } \right] \Delta \Dgrad \bigg|_- - \mbf{v}_0 \bigg|_+ \bigg] \wedge \mbf{n}_0 \cdot \delta \pot ds_0 \nonumber \\
&+ \int\limits_{\partial \vol_0} \bigg[ \mbf{T} \mbf{n}_0 \cdot \delta \rchi + \mbf{v}_0 \wedge \mbf{n}_0 \cdot \delta \pot \bigg] \, ds_0 
= 0.
\end{align}

Since the variations $\delta \rchi$ and $\delta \pot$ are arbitrary, we arrive at the equations \eqref{eqn: second variation based PDE}.

\bibliographystyle{./author-year-prashant}
\bibliography{./references_used2}

\begin{thebibliography}{34}
\newcommand{\enquote}[1]{``#1''}
\providecommand{\natexlab}[1]{#1}

\bibitem[{Bertoldi and Gei(2011)}]{Bertoldi2011}
Bertoldi K. and Gei M.
\newblock \enquote{{Instabilities in multilayered soft dielectrics}}.
\newblock \emph{Journal of the Mechanics and Physics of Solids}, 59(1):18--42
  (2011)

\bibitem[{Bustamante et~al.(2009{\natexlab{a}})Bustamante, Dorfmann, and
  Ogden}]{Bustamante2009}
Bustamante R., Dorfmann A., and Ogden R.W.
\newblock \enquote{{Nonlinear electroelastostatics: a variational framework}}.
\newblock \emph{Zeitschrift f{\"{u}}r Angewandte Mathematik und Physik},
  60:154--177 (2009{\natexlab{a}})

\bibitem[{Bustamante et~al.(2009{\natexlab{b}})Bustamante, Dorfmann, and
  Ogden}]{Bustamante2009b}
Bustamante R., Dorfmann A., and Ogden R.W.
\newblock \enquote{{On electric body forces and Maxwell stresses in nonlinearly
  electroelastic solids}}.
\newblock \emph{International Journal of Engineering Science},
  47(11-12):1131--1141 (2009{\natexlab{b}})

\bibitem[{Dorfmann and Ogden(2005)}]{Dorfmann2005a}
Dorfmann A. and Ogden R.W.
\newblock \enquote{{Nonlinear electroelasticity}}.
\newblock \emph{Acta Mechanica}, 174(3-4):167--183 (2005)

\bibitem[{Dorfmann and Ogden(2006)}]{Dorfmann2006}
Dorfmann A. and Ogden R.W.
\newblock \enquote{{Nonlinear electroelastic deformations}}.
\newblock \emph{Journal of Elasticity}, 82(2):99--127 (2006)

\bibitem[{Dorfmann and Ogden(2010)}]{Dorfmann2010a}
Dorfmann A. and Ogden R.W.
\newblock \enquote{{Electroelastic waves in a finitely deformed electroactive
  material}}.
\newblock \emph{IMA Journal of Applied Mathematics}, 75(4):603--636 (2010)

\bibitem[{Dorfmann and Ogden(2014{\natexlab{a}})}]{Dorfmann2014a}
Dorfmann L. and Ogden R.W.
\newblock \enquote{{Instabilities of an electroelastic plate}}.
\newblock \emph{International Journal of Engineering Science}, 77:79--101
  (2014{\natexlab{a}})

\bibitem[{Dorfmann and Ogden(2014{\natexlab{b}})}]{Dorfmann2014b}
Dorfmann L. and Ogden R.W.
\newblock \emph{{Nonlinear theory of electroelastic and magnetoelastic
  interactions}}.
\newblock Springer (2014{\natexlab{b}})

\bibitem[{Ericksen(2007)}]{Ericksen2007a}
Ericksen J.L.
\newblock \enquote{{Theory of elastic dielectrics revisited}}.
\newblock \emph{Archive for Rational Mechanics and Analysis}, 183(2):299--313
  (2007)

\bibitem[{Gelfand and Fomin(2003)}]{Gelfand2003}
Gelfand I.M. and Fomin S.V.
\newblock \emph{{Calculus of Variations}}.
\newblock Dover Publications (2003)

\bibitem[{Giaquinta and Hildebrandt(2010)}]{Giaquinta2010}
Giaquinta M. and Hildebrandt S.
\newblock \emph{{Calculus of Variations I}}.
\newblock Springer (2010)

\bibitem[{Gurtin(1981)}]{Gurtin1981}
Gurtin M.E.
\newblock \emph{{An Introduction to Continuum Mechanics}}.
\newblock Academic Press (1981)

\bibitem[{Hill(1957)}]{Hill1957}
Hill R.
\newblock \enquote{{On Uniqueness and Stability in the theory of finite elastic
  strain}}.
\newblock \emph{Journal of the Mechanics and Physics of Solids}, 5:229--241
  (1957)

\bibitem[{Itskov(2018)}]{Itskov2018}
Itskov M.
\newblock \emph{{Tensor Algebra and Tensor Analysis for Engineers: With
  Applications to Continuum Mechanics, 5th edition}}.
\newblock Springer (2018)

\bibitem[{Jung et~al.(2008)Jung, Kim, and Choi}]{Jung2008}
Jung K., Kim K.J., and Choi H.R.
\newblock \enquote{{A self-sensing dielectric elastomer actuator}}.
\newblock \emph{Sensors and Actuators A}, 143:343--351 (2008)

\bibitem[{Knowles(1997)}]{Knowles1997}
Knowles J.K.
\newblock \emph{{Linear Vector Spaces and Cartesian Tensors}}.
\newblock OUP USA (1997)

\bibitem[{Kofod(2001)}]{Kofod2001}
Kofod G.
\newblock \emph{{Dielectric elastomer actuators}}.
\newblock Ph.D. thesis, Technical University of Denmark (2001)

\bibitem[{Koiter(1965)}]{Koiter2}
Koiter W.
\newblock \enquote{The energy criterion of stability for continuous bodies}.
\newblock \emph{Proceedings, Koninklijke Nederlandse Akademie van
  Wettenschappen}, 68:178--202 (1965).
\newblock Amsterdam, Series B, Phys. Sciences

\bibitem[{Koiter(1970)}]{Koiter}
Koiter W.T.
\newblock \enquote{The stability of elastic equilibrium}.
\newblock Technical report, DTIC Document (1970)

\bibitem[{Liu(2014)}]{Liu2014}
Liu L.
\newblock \enquote{{An energy formulation of continuum
  magneto-electro-elasticity with applications}}.
\newblock \emph{Journal of the Mechanics and Physics of Solids}, 63:451--480
  (2014)

\bibitem[{McMeeking and Landis(2005)}]{McMeeking2005}
McMeeking R.M. and Landis C.M.
\newblock \enquote{{Electrostatic forces and stored energy for deformable
  dielectric materials}}.
\newblock \emph{Journal of Applied Mechanics}, 72(4):581--590 (2005)

\bibitem[{Michel et~al.(2008)Michel, Bormann, Jordi, and Fink}]{Michel2008}
Michel S., Bormann A., Jordi C., and Fink E.
\newblock \enquote{{Feasibility studies for a bionic propulsion system of a
  blimp based on dielectric elastomers}}.
\newblock \emph{Proceedings of SPIE - EAPAD}, 4332:1--15 (2008)

\bibitem[{O'Halloran et~al.(2008)O'Halloran, O'Malley, and
  McHugh}]{O'Halloran2008}
O'Halloran A., O'Malley F., and McHugh P.
\newblock \enquote{{A review on dielectric elastomer actuators, technology,
  applications, and challenges}}.
\newblock \emph{Journal of Applied Physics}, 104(7):71101--71110 (2008)

\bibitem[{Ozsecen et~al.(2010)Ozsecen, Sivak, and Mavroidis}]{Ozsecen2010}
Ozsecen M.Y., Sivak M., and Mavroidis C.
\newblock \enquote{{Haptic interfaces using dielectric electroactive
  polymers}}.
\newblock \emph{In} M.~Tomizuka, C.B. Yun, V.~Giurgiutiu, and J.P. Lynch
  (Eds.), \enquote{Proceedings of SPIE - Sensors and Smart Structures
  Technologies for Civil, Mechanical, and Aerospace Systems}, page 7647 (2010)

\bibitem[{Pak and Herrmann(1986)}]{Pak1986}
Pak Y.E. and Herrmann G.
\newblock \enquote{{Conservation laws and the material momentum tensor for the
  elastic dielectric}}.
\newblock \emph{International Journal of Engineering Science}, 24:1365--1374
  (1986)

\bibitem[{Pelrine et~al.(2001)Pelrine, Kornbluh, Eckerle, Jeuck, Oh, Pei, and
  Stanford}]{Pelrine2001}
Pelrine R., Kornbluh R., Eckerle J., Jeuck P., Oh S., Pei Q., and Stanford S.
\newblock \enquote{{Dielectric elastomers: generator mode fundamentals and
  applications}}.
\newblock \emph{Proceedings of SPIE - Smart Structures and Materials},
  4329:148--156 (2001)

\bibitem[{Pelrine et~al.(2000)Pelrine, Kornbluh, Pei, and
  Joseph}]{Pelrine2000a}
Pelrine R., Kornbluh R., Pei Q., and Joseph J.
\newblock \enquote{{High-Speed Electrically Actuated Elastomers with Strain
  Greater Than 100{\%}}}.
\newblock \emph{Science}, 287(5454):836--839 (2000)

\bibitem[{Saxena et~al.(2014)Saxena, Vu, and Steinmann}]{Saxena2014a}
Saxena P., Vu D.K., and Steinmann P.
\newblock \enquote{{On rate-dependent dissipation effects in
  electro-elasticity}}.
\newblock \emph{International Journal of Non-Linear Mechanics}, 62:1--11 (2014)

\bibitem[{Shintake et~al.(2016)Shintake, Rosset, Schubert, Floreano, and
  Shea}]{Shintake2016}
Shintake J., Rosset S., Schubert B., Floreano D., and Shea H.
\newblock \enquote{{Versatile Soft Grippers with Intrinsic Electroadhesion
  Based on Multifunctional Polymer Actuators}}.
\newblock \emph{Advanced Materials}, 28(2):231--238 (2016)

\bibitem[{Toupin(1956)}]{Toupin1956}
Toupin R.
\newblock \enquote{{The elastic dielectric}}.
\newblock \emph{Journal of Rational Mechanics and Analysis}, 5(6):849--915
  (1956)

\bibitem[{van~der Heijden(2009)}]{Heijden}
van~der Heijden A.M.A. (Ed.).
\newblock \emph{W. T. Koiter's Elastic Stability of Solids and Structures}.
\newblock Cambridge University Press (2009)

\bibitem[{Vu and Steinmann(2012)}]{Vu2012a}
Vu D.K. and Steinmann P.
\newblock \enquote{{On the spatial and material motion problems in nonlinear
  electro-elastostatics with consideration of free space}}.
\newblock \emph{Mathematics and Mechanics of Solids}, 17(8):803--823 (2012)

\bibitem[{Wingert et~al.(2006)Wingert, Lichter, and Dubowsky}]{Wingert2006}
Wingert A., Lichter M.D., and Dubowsky S.
\newblock \enquote{{On the design of large degree-of-freedom digital
  mechatronic devices based on bistable dielectric elastomer actuators}}.
\newblock \emph{IEEE/ASME Transactions on Mechatronics}, 11(4):448--456 (2006)

\bibitem[{Yang and Batra(1995)}]{Yang1995}
Yang J. and Batra R.C.
\newblock \enquote{{Mixed variational principles in non-linear
  electroelasticity}}.
\newblock \emph{International Journal of Non-Linear Mechanics}, 30:719--725
  (1995)

\end{thebibliography}
\end{document}